\documentclass[12pt]{article}
\usepackage[T1]{fontenc}
\usepackage[latin9]{inputenc}
\usepackage{geometry}
\usepackage{bm}
\usepackage{float}
\usepackage{graphicx}

\usepackage{amsfonts}
\usepackage{amsmath}
\usepackage{amssymb}
\usepackage{amsthm}
\usepackage{setspace}
\usepackage{booktabs}
\usepackage[normalem]{ulem}
\usepackage{xcolor}
\usepackage[backend=biber,style=apa, doi=false,isbn=false,url=false,eprint=false]{biblatex}
\usepackage{authblk}
\usepackage{comment}
\usepackage[medium]{titlesec}
\usepackage{graphicx,caption}

\DeclareMathOperator*{\argmax}{arg\,max}

\definecolor{darkred}{RGB}{150,50,50}
\definecolor{brown}{RGB}{250,100,100}
\definecolor{green}{RGB}{000,150,100}
\definecolor{purple}{RGB}{250,000,180}

\newtheorem{remark}{Remark}
\newtheorem{thm}{Theorem}

\def\muhat{\widehat{\mu}}
\def\PPVhat{\widehat{\PPV}}
\def\NPVhat{\widehat{\NPV}}
\def\ROChat{\widehat{\ROC}}
\def\AUChat{\widehat{\AUC}}
\def\TPRhat{\widehat{\TPR}}
\def\FPRhat{\widehat{\FPR}}
\def\supwgt{^{\scriptscriptstyle \sf wgt}}
\def\TPR{\rm{TPR}}
\def\FPR{\rm{FPR}}
\def\ROC{\rm{ROC}}
\def\PPV{\rm{PPV}}
\def\NPV{\rm{NPV}}
\def\AUC{\rm{AUC}}
\def\bbeta{\bm\beta}
\def\balpabar{\bm{\bar\alpha}}
\def\balphatilde{\bm{\widetilde\alpha}}
\def\bbetatilde{\bm{\widetilde\beta}}
\def\betatilde{\widetilde\beta}
\def\bbetahat{\bm{\widehat\beta}}
\def\bbetabar{\bm{\bar\beta}}
\def\balpha{\bm\alpha}
\def\balphahat{\bm{\widehat\alpha}}
\def\balphabar{\bm{\bar\alpha}}
\def\bX{\bm X}
\def\bx{\bm x}
\def\IL{\mathcal I_L}
\def\IU{\mathcal I_U}
\def\IT{\mathcal I_T}
\def\pto{\overset{P}\to}
\def\nhalf{n^{1/2}}

\def\Isc{\mathcal{I}}
\def\Psc{\mathcal{P}}
\def\Tsc{\mathcal{T}}
\def\Ptbetabar{ {\mathcal P}_{0}\supbetabar}
\def\Ptbetabari{ {\mathcal P}_{0,i}\supbetabar}
\def\Ptbetabarj{ {\mathcal P}_{0,j}\supbetabar}
\def\Ptbetahat{ {\mathcal P}_{0}\supbetahat}
\def\Ptbetahati{ {\mathcal P}_{0,i}\supbetahat}
\def\Pthatbetahat{\widehat{\Psc}_{0}\supbetahat}
\def\Pthatbetahatstar{\widehat{\Psc}_{0}\supbetahatstar}
\def\Pthatbetahati{\widehat{\Psc}_{0,i}\supbetahat}
\def\Pthatbetahatstari{\widehat{\Psc}_{0,i}\supbetahatstar}
\def\Pthatbetahatnki{\widehat{\Psc}_{0,i}\supbetahatnk}
\def\Pthatbetabar{\widehat{\Psc}_{0}\supbetabar}
\def\Pthatbetabari{\widehat{\Psc}_{0,i}\supbetabar}
\def\Ptbeta{\Psc_{0}\supbeta}
\def\Ptbetai{\Psc_{0,i}\supbeta}
\def\mtbetabar{m_{0}\supbetabar}
\def\mthatbetahat{\mthat\supbetahat}
\def\mthatbetabar{\mthat\supbetabar}
\def\mthatbetahatstar{\mthat\supbetahatstar}
\def\mthat{\widehat{m}_0}
\def\Ftbetabar{F_{0}\supbetabar}
\def\Fhat{\widehat{F}}
\def\Fthatbeta{\Fhat_{0}\supbeta}
\def\Fthatbetahat{\Fhat_{0}\supbetahat}
\def\Ebb{\mathbb{E}}
\def\Pbb{\mathbb{P}}
\def\Ebbt{\mathbb{E}_0}
\def\Pbbt{\mathbb{P}_0}
\def\sumiNt{\sum_{i=1}^{N_t}}
\def\sumjNt{\sum_{j=1}^{N_t}}
\def\supSL{^{\scriptscriptstyle \sf SL}}
\def\supSTEAM{^{\scriptscriptstyle \sf STEAM}}
\def\supSTEAMstar{^{\scriptscriptstyle \sf STEAM*}}
\def\supSTEAMk{^{\scriptscriptstyle \sf STEAM,(k)}}
\def\supSTEAMinv{^{\scriptscriptstyle \sf STEAM \scriptstyle -1}}
\def\wtilde{\widetilde{w}}
\def\cbar{\bar{c}}
\def\chat{\widehat{c}}
\def\subut{_{u_0}}
\def\subbeta{_{\scriptscriptstyle \bbeta}}
\def\subbetahat{_{\scriptscriptstyle \bbetahat}}
\def\subbetabar{_{\scriptscriptstyle \bbetabar}}
\def\supbeta{^{\scriptscriptstyle \bbeta}}
\def\supbetahat{^{\scriptscriptstyle \bbetahat}}
\def\supbetahatstar{^{\scriptscriptstyle \bbetahat^\ast}}
\def\supbetabar{^{\scriptscriptstyle \bbetabar}}
\def\supbetahatnk{^{\scriptscriptstyle \bbetahat^{(-k)}}}
\def\supalphahat{^{\scriptscriptstyle \balphahat}}
\def\supalpha{^{\scriptscriptstyle \balpha}}
\def\supalphabar{^{\scriptscriptstyle \balphabar}}
\def\bX{\bm X}
\def\bx{\bm x}
\def\IL{\mathcal I_L}
\def\IU{\mathcal I_U}
\def\IT{\mathcal I_T}
\def\pto{\overset{P}\to}
\def\nhalf{n^{1/2}}
\def\nnhalf{n^{-1/2}}
\def\Psc{\mathcal{P}}
\def\Psct{\mathcal{P}_0}

\def\Psctbbetahat{\Psct\supbetahat}
\def\trans{^{\scriptscriptstyle \sf T}}
\def\pihat {\widehat \pi}
\def\what {\widehat w}
\def\wbar {\bar w}
\def\xihat{\widehat{\xi}}
\def\bC{\mathbf{C}}
\def\bphi{\boldsymbol{\phi}}

\title{\Large Semi-supervised Transfer Learning for Evaluation of Model Classification Performance}
\author[1]{\normalsize Linshanshan Wang}
\author[1]{\normalsize Xuan Wang}
\author[2]{\normalsize Katherine P. Liao} 
\author[1,3]{\normalsize Tianxi Cai}
\affil[1]{\footnotesize Department of Biostatistics, Harvard T.H. Chan School of Public Health}
\affil[2]{\footnotesize Division of Rheumatology, Brigham and Women's Hospital}
\affil[3]{\footnotesize Department of Biomedical Informatics, Harvard Medical School}
\date{\vspace{-5ex}}

\begin{document}
\maketitle
\newrefsection
\begin{abstract}
\singlespacing
    In many modern machine learning applications, frequent encounters of covariate shift and label scarcity have posed challenges to robust model training and evaluation. Numerous transfer learning methods have been developed to robustly adapt the model itself to some unlabeled target populations using existing labeled data in a source population. However, there is a paucity of literature on transferring performance metrics of a trained model. In this paper, we aim to evaluate the performance of a trained binary classifier on unlabeled target population based on receiver operating characteristic (ROC) analysis. We proposed {\bf S}emi-supervised {\bf T}ransfer l{\bf E}arning of {\bf A}ccuracy {\bf M}easures (STEAM), an efficient three-step estimation procedure that employs 1) double-index modeling to construct calibrated density ratio weights and 2) robust imputation to leverage the large amount of unlabeled data to improve estimation efficiency. We establish the consistency and asymptotic normality of the proposed estimator under correct specification of either the density ratio model or the outcome model. We also correct for potential overfitting bias in the estimators in finite samples with cross-validation. We compare our proposed estimators to existing methods and show reductions in bias and gains in efficiency through simulations. We illustrate the practical utility of the proposed method on evaluating prediction performance of a phenotyping model for Rheumatoid Arthritis (RA) on a temporally evolving EHR cohort.
\end{abstract}
\begin{singlespace*}
\end{singlespace*}

\section{Introduction}
The increasing adoption of electronic health records (EHR) for clinical care has enabled many opportunities for translational and clinical research (\cite{hripcsak_nextgeneration_2013}, \cite{miotto_deep_2016}). However, a major limitation of EHR data is the lack of precise information on disease phenotypes. Readily available EHR features such as diagnosis codes often exhibit poor specificity that can bias or depower the downstream study (\cite{liao_electronic_2010}, \cite{cipparone_inaccuracy_2015}). Meanwhile, manual annotation of phenotypes via chart review is laborious and unscalable. To overcome these challenges, numerous supervised phenotyping classification algorithms have been developed for prediction of disease status of individual patients (\cite{carroll_portability_2012}, \cite{xia_modeling_2013}, \cite{liao_development_2015}, \cite{liao_methods_2015}).

To build a supervised phenotyping model, researchers often manually annotate outcome information for a randomly chosen subset of the large EHR dataset, and then use these labeled observations for model training and validation. Once a model is trained and validated in one population, it is often then used for other populations such as a similar population from a different healthcare system (\cite{carroll_portability_2012}) or the same EHR cohort updated over time (\cite{huang_impact_2020}). Generalizing a classification model trained in a \textit{source} population to a \textit{target} population, however, requires caution due to the shift in the covariate distribution owing to the heterogeneity in the underlying patient populations. It is well-known that such shifts in the distribution of the covariates, often referred to as \textit{covariate shift}, may have a large impact on the performance of a prediction algorithm trained in source cohort on target cohort (\cite{rasmy_study_2018}). It is thus important to assess the portability of the algorithm in face of such changes in datasets before its usage in the target sites.

In these scenarios, methods to accurately evaluate the the performance of these algorithms in target cohort is especially valuable. If outcome information is available for some patients in target cohort, for instance, via chart review, then the standard methods can be applied to estimate the prediction performance parameters on target dataset. Recently, robust semi-supervised methods have also been proposed to improve the estimation efficiency for prediction performance measures (\cite{gronsbell_semi_2018}). The issue, however, is that typically, performing chart review for every new target population is infeasible due to time and resource constraints. It is thus highly desirable to robustly and efficiently estimate the prediction performance parameters in target cohort without requiring any outcome information in target dataset.

Although the problem of covariate shift has been extensively studied in recent literature of transfer learning, most existing work focused on robust adaptation of the model itself to the target population (\cite{chen_robust_2016}, \cite{wen_robust_2014}, \cite{rotnitzky_improved_2012}, \cite{reddi_doubly_2015}, \cite{liu_doubly_2020}). There is a paucity of literature on transferring performance metrics of a trained model based on the commonly used ROC analysis (\cite{pepe_statistical_2003}). \textcite{inoue_appestimation_2018} used density ratio weighting methods to assess model performance in the target population. However, their estimator do not protect against the mis-specification of the density ratio model. \textcite{xu_estimation_2022} proposed a method based on parametric bootstrap which does not require estimation of the density ratio model. However, their focus was on the out-of-sample error of the prediction model, not estimating the accuracy meausures based on the ROC curve. Moreover, their method does not protect against mis-specification of the prediction model. Doubly robust methods for estimating accuracy measures have been proposed in verification bias literature (\cite{alonzo_assessing_2005}, \cite{rotnitzky_doubly_2006}, \cite{fluss_estimation_2009}), and recently in the context of estimating the AUC in a target population that has a different data distribution compared to the source population (\cite{li_estimating_2022}). Nevertheless, no existing method leveraged both unlabeled source and target data to improve estimation efficiency. To address this gap in methodology, we propose a {\bf S}emi-supervised {\bf T}ransfer l{\bf E}arning of {\bf A}ccuracy {\bf M}easures (STEAM) in the unlabeled target cohort. We employ a double-index modeling approach to construct calibrated density ratio weights similar to \textcite{cheng_estimating_2020}, incorporating both the covariate shift model and the outcome imputation model. Although also taking a re-weighting form, the STEAM estimators are doubly robust with consistency attained when either the density ratio model or the classification model is correctly specified. The large amount of unlabeled source and target data enabled us to efficiently estimate the density ratio and the imputation model. 

The rest of the paper is organized as follows: We formalize the problem of interest in Section 2. In section 3, we detail the STEAM estimation procedure along with a perturbation resampling procedure for inference. We conduct simulation studies to evaluate the
finite sample robustness and efficiency of the proposed method in Section 4. We then illustrate the practical utility of the proposed method on evaluating prediction performance of a phenotyping model for Rheumatoid Arthritis (RA) on temporally evolving EHR cohorts in Section 5.

\section{Problem Setup}
\subsection{Notations and Assumptions}

Let $Y$ denote the binary phenotype of interest, and $\bX=(1, X_1, \cdots, X_p)\trans $ the $p+1$-dimensional covariate vector for some fixed $p$. 
The source data, indexed by $S=1$, consist of $n$ independent and identically distributed (iid) labelled data $\mathcal L = \{(\bX_i, Y_i): i\in\mathcal I_L\}$ where $n=|\mathcal I_L|$, and $N$ iid unlabelled data $\mathcal U = \{\bX_i: i\in\mathcal I_U\}$ where $N=|\mathcal I_U|$. The target data, indexed by $S=0$, consists of $N_t$ iid unlabelled data $\mathcal T = \{\bX_i: i\in\IT\}$, where $N_t=|\mathcal I_T|$. The entire observed data $\mathcal D = \mathcal L\cup \mathcal U\cup\mathcal T$ can be written as $\{(S_iL_iY_i, \bX_i, S_i): i=1,2,\dots,N \}$, where $L$ is an indicator of labelling. We assume that $N/N_t \rightarrow c \in (0, \infty)$, $\log(N)/\log(n) \to r_0 > \frac{3}{2}$ such that $\nhalf N^{-1/3} \to 0$, as $n\rightarrow \infty$. 
We assume random labeling in source population and hence $L \perp (\bX, Y)$. Furthermore, we require the standard covariate shift assumption with
$p(y,\bx|s)= p_s(\bx)q(y|\bx)$, where $p(y,\bx|s)$ is the joint density of $Y,\bX$ evaluated at $(y, \bx)$ given $S=s$, $p_s(\bx)$ denotes the probability density of $\bX$ at $\bx$ given $S=s$ and $q(y|\bx)$ denotes the conditional density of $Y$ at $y$ given $\bX=\bx$, which is shared between the source and target populations. Throughout, we let $\Pbb_s$ an $\Ebb_s$ denote the probability distribution and expectation taken over the population $S = s$. 

\subsection{Classification Rule}
The aim is to assess the accuracy of a binary classification rule for $Y$ given $\bX$ developed using the source data. Although our STEAM approach can be easily extended to other regression models, for the ease of presentation, we focus on the {\em working} generalized linear model (GLM)
\begin{equation}\label{eq:mu}
    P(Y=1|\bX, L=1)=g(\bm{\beta\trans \bX}),\ \ \bbeta=(\beta_0, \beta_1, \cdots, \beta_p)\trans , 
\end{equation}
where $g$ is a specified smooth link function such that $\ell_g(\bbeta; y,\bx)=y \log g(\bbeta\trans\bx) + (1-y)\log \{1- g(\bbeta\trans\bx))\}$ is convex in $\bbeta$. One can also easily accommodate non-linear effects by including non-linear basis functions of $\bX$. We consider standard estimation procedures for $\bbeta$. As an example, one may employ an adaptive LASSO penalized estimator (\cite{zou_adaptive_2006}) to stabilize the finite sample estimation when $p$ is not very small relative to $n$:
\begin{equation}\label{eq:beta}
    \bbetahat = \argmax_{\bm\beta} \left\{n^{-1} \sum_{i\in\IL} \ell_g(\bbeta;Y_i,\bX_i) - \lambda_{\mu,n}  \|(\bbeta_{-1}/\bbetatilde_{-1})^\gamma\|_1 \right\}
\end{equation}
where $\bbetatilde = (\betatilde_0,
\betatilde_1,..., \betatilde_{p})\trans$ is a root-$n$ consistent initial estimate of $\bbeta$, $\lambda_{\mu,n}$ is a tuning parameter such that $\nhalf\lambda_{\mu,n}\to 0$ and $n^{(1-\nu)(1+\gamma)/2}\lambda_{\mu,n}\to\infty$ with $\gamma>2\nu(1-\nu)$ (\cite{zou_adaptive_2009}).
Following results given in \textcite{zou_adaptive_2009}, $\bbetahat$ is a root-$n$ consistent for the population parameter $$\bbetabar = \argmax E\{\ell_g(\bbeta; Y_i, \bX_i) \mid S_i = 1\}$$ and attains oracle property under sparsity of $\bbetabar$, regardless if the fitted model (\ref{eq:mu}) holds. Our proposed STEAM procedure is not restricted to any specific estimation procedure provided that $\bbetahat$ is a regular root-$n$ estimators for $\bbetabar$. 


\subsection{Accuracy Measures of interest}
With a given $\bbetabar$ trained by source data, we may classify subjects in the target population with the highest scores of $\bbetabar\trans\bX$, i.e.  $\Ptbetabar \equiv \Ftbetabar(\bbetabar\trans \bX)\ge c$ for some fraction $c$, as having the phenotype of interest, where $\Ftbetabar (a)=\Pbb_0(\bbetabar\trans \bX \le a)$.
To identify a desirable threshold  $c$ and evaluate the classification accuracy in the unlabeled target population, we consider the commonly employed receiver operating characteristic (ROC) analysis  (\cite{pepe_statistical_2003}). Specifically, the true positive rate (TPR) and false positive rate (FPR)  of the binary classification rule $I(\Ptbetabar \ge c)$ in the target population are defined as
$$
\begin{aligned}
\TPR_0(c) &= \Pbb_0(\Ptbetabar\ge c | Y=1) , \quad 
\FPR_0(c) &= \Pbb_0(\Ptbetabar\ge c | Y=0) 
\end{aligned}
$$
The ROC curve, $\ROC_0 = \TPR_0\{\FPR_0^{-1}(u) \}$, summarizes the trade-off between the TPR and FPR functions. 
The area under the ROC curve, $\AUC_0=\int_0^1 \ROC_0(u)du$, summarizes the overall prediction performance of $\mathcal P_{\hat\beta}$. Additionally, a threshold value $c_0$ for classifying an individual as having the phenotype, namely $\Ptbetabar \ge c_0$, is often chosen to achieve a desired FPR level $u_0$. Once the threshold value $c_0$ is determined, one may summarize the predictive performance of  $I(\Ptbetabar\ge c_0)$, based on positive predictive value (PPV) and negative predictive value (NPV) defined as
$$
\begin{aligned}
\PPV_0(c_0) &= \Pbb_0(Y=1|\Ptbetabar \ge c_0) , \quad 
\NPV_0(c_0) &= \Pbb_0(Y=0|\Ptbetabar < c_0) 
\end{aligned}
$$
Directly estimating theses accuracy parameters using the source data will result in inconsistency due to the covariate shift as well as potential model mis-specification of the model for $P(Y=1|\bX)$ (\cite{liu_double_2021}). To correct for the bias, it is natural to incorporate importance sampling weighting. For example, with $\bbetabar$ estimated as $\bbetahat$, one may estimate $\TPR_0$ as 
\begin{equation}\label{eq:wgt}
    \TPRhat_0\supwgt(c) = \frac{\sum_{i\in\IL}  I(\Pthatbetahati \ge c) \widehat w(\bX_i)\cdot Y_i }{\sum_{i\in\IL} \widehat w(\bX_i)\cdot Y_i }
\end{equation}
where $\widehat w(\bX)$ is an estimate for the density ratio $$w(\bX)=\Pbb_0(\bX)/\Pbb_1(\bX)=\{1-\pi(\bX)\}/\pi(\bX)\cdot P(S=1)/P(S=0),\quad \pi(\bX)=P(S=1|\bX),$$ $\Pthatbetahati = \Fthatbetahat(\bbetahat\trans\bX_i)$, and $\Fthatbeta(a) = N_t^{-1}\sum_{i\in \Isc_T} I(\bbeta\trans\bX_i \le a)$. However, the validity of this re-weighting estimator heavily relies on the consistency of $\widehat w(\bX)$ for $w(\bX)$ and can perform poorly when the density ratio model is not well estimated. Moreover, the estimator suffers from low precision when $n$ is small, which is often the case in EHR settings. To improve the robustness and efficiency of the estimation for the accuracy parameters, we next propose the STEAM estimators that leverage the unlabeled data.

\section{Methods}
\subsection{Estimation Procedure}
To motivate the STEAM estimation procedure, we first note that 
$$
\TPR_0(c) = \frac{\Ebbt\{Y\cdot I(\Ptbetabar\ge c)\}}{\Ebbt(Y)} = \frac{\Ebbt\{ \mtbetabar(\Ptbetabar)\cdot I(\Ptbetabar\ge c)\}}{\Ebbt\{\mtbetabar(\Ptbetabar)\}} = \frac{\Ebbt\{ \int_c^1 \mtbetabar(q) dq\} }{\Ebbt\{ \int_0^1 \mtbetabar(q) dq\}}
$$
where $\mtbetabar(q) = \Pbb_0(Y=1|\Ptbetabar=q)$. Even though $\Pbb_0(Y=1\mid \bX) = \Pbb_1(Y=1\mid \bX)$, $\Pbb_0(Y=1\mid \Ptbetabar) \ne \Pbb_1(Y=1\mid \Ptbetabar)$ under covariate shift and possible mis-specification of the outcome model (\ref{eq:mu}). 
To enable robust transfer of knowledge from the source to target while leveraging unlabeled data, the STEAM estimation procedure consists of three key steps: 
(I) construct calibrated density ratio weight $\what_i$ via double index modeling; (II) derive non-parametric calibrated estimator of $\mtbetabar(c)$; and (III) obtain the final STEAM estimator by projecting to the unlabeled target data using the estimated $\mtbetabar(\cdot)$.

\def\bPsi{\bm{\Psi}}
\def\bh{\bm{h}}

\paragraph{Step I: Calibrated Density Ratio Weight Estimation}
The calibrated density ratio weight estimator involves both an initial estimator for $\pi(\bX)=P(S=1\mid\bX)$ and the estimated outcome model. To estimate $\pi(\bX)$, we consider a parametric working model, 
\begin{equation}\label{eq:pi}
P(S=1|\bX)=g_\pi(\balpha\trans \bPsi),  
\end{equation}
where $g_\pi$ is a known link function, $\bPsi$ is a $q$-dimensional vector of transformed $\bX$ that may include non-linear bases of $\bX$ with the first element being 1,  and $\bm\alpha$ is the 
unknown parameter.  Since $\pi(\cdot)$ doesn't involve the outcome, we may use the unlabeled data to obtain a regularized estimator for $\balphabar = \argmax_{\balpha} \Ebb\{ \ell_{g_\pi}(\balpha;S_i,\bX_i) \}$:
$$
\balphahat = \argmax_{\balpha} \left\{(N+N_t)^{-1} \sum_{i\in\IU\cup\IT} \ell_{g_\pi}(\balpha;S_i,\bPsi_i) - \lambda_{\pi,N} \|\balpha_{-1}/\balphatilde_{-1}\|_1 \right\} 
$$
with properly chosen tuning parameters $\lambda_{\pi,N}$, $\gamma$, and an initial estimator $\balphatilde$ for $\balphabar$, where the expectation is taken over the pooled source and target population. 
Similar to the outcome model, one may show that $\balphahat$ is a regular estimator for $\balphabar$.  
We then obtain a non-parametrically calibrated estimate of $\pi(\bX)$ by smoothing $S$ over the two scores $\balphabar\trans\bX$ and $\bbetabar\trans\bX$. Specifically, we estimate $\pi(\bX)$ as 
$\pihat(\bX; \balphahat, \bbetahat)$, 
\begin{align*}
    \pihat(\bX; \balphahat, \bbetahat) &= \frac{\sum_{i\in \IU \cup \IT} K_{\pi,h_1}\left\{(\balphahat, \bbetahat)\trans  (\bX_i-\bX)\right\}S_i}{\sum_{i\in \IU \cup \IT}  K_{\pi,h_1}\left\{(\balphahat, \bbetahat)\trans  (\bX_i-\bX)\right\}} 
\end{align*}
and then construct the calibrated density ratio weight for the $i$th subject with $\bX_i$ as 
$$
\pihat_i \equiv \what(\bX_i; \balphahat, \bbetahat) = \frac{1-\pihat(\bX_i; \balphahat, \bbetahat)}{\pihat(\bX_i; \balphahat, \bbetahat)} ,
$$
where $K_{\pi, h_1}\{(u_1,u_2)\trans\} = h_1^{-2}K(u_1/h_1)K(u_2/h_1)$, $K(\cdot)$ is a smooth symmetric density function, and $h_1=O(N^{-1/6})$. 

\paragraph{Step II: Calibrated Estimate of $\mtbetabar(q) = \Pbb_0(Y=1|\Ptbetabar=q)$.}
In step II, we obtain a non-parametrically calibrated estimate of the conditional risk function $\mtbetabar(q)$ via kernel smoothing as
$$
\mthatbetahat(q) = \frac{\sum_{i\in\IL}K_{m,h_2}( \Pthatbetahati-q)\what_iY_i}{\sum_{i\in\IL}K_{m.h_2}(\Pthatbetahati-q)\what_i} 
$$
where $\Pthatbetahati = \Pthatbetahat(\bX_i)$, $\Pthatbetahat(\bx)=\widehat F_{\bbetahat} (\bbetahat\trans \bx)$,  $\widehat F_{\bbeta}(t) = N_t^{-1}\sum_{i\in \IT} I(\bbeta\trans \bX_i\le t)$, $K_{m,h_2}(u)=h_2^{-1}K(u/h_2)$, and $h_2$ is a bandwidth such that $nh_2^2\to\infty$ and $nh_2^4\to0$ as $n\to\infty$. 

\paragraph{Step III: Projection to $\Tsc$ with $\mthatbetahat(\cdot)$} Finally, we construct a plug-in STEAM estimator by projecting to $\Tsc$ using the imputed outcomes $\{\mthatbetahat(\Pthatbetahati), i \in \IT\}$. Specifically, the STEAM estimator for $\TPR_0(c)$ is obtained as
$$
  \TPRhat\supSTEAM_0(c) = \frac{\sum_{i\in\IT}  I(\Pthatbetahati \ge c) \mthatbetahat(\Pthatbetahati)}{\sum_{i\in\IT}  \mthatbetahat(\Pthatbetahati)}  
$$

\begin{remark}
The calibrated estimator $\mthatbetahat(q)$ consistently estimates $\mtbetabar(q)$ if either \eqref{eq:mu} or \eqref{eq:pi} is correct. The STEAM approach to imputation is appealing as it achieves double robustness by leveraging the large unlabeled data and only requiring one-dimensional smoothing over the small labeled data.  
\end{remark}

Similarly we may construct a STEAM estimator of $\FPR_0(c)$ as
$$
\FPRhat\supSTEAM_0(c) = \frac{\sum_{i\in\IT}  I(\Pthatbetahati \ge c) \{ 1-\mthatbetahat(\Pthatbetahati)\}}{\sum_{i\in\IT}  \{1-\mthatbetahat(\Pthatbetahati)\}} $$
The STEAM estimates of the ROC curve and the resulting AUC are
$$
\ROChat\supSTEAM_0 = \TPRhat\supSTEAM_0\{ {{\FPRhat\supSTEAMinv}_0}(u) \}, \quad \mbox{and}\quad \AUChat\supSTEAM_0=\int_0^1 \ROChat\supSTEAM_0(u)du.$$ 
Furthermore, when a threshold value $\cbar\subut$ is chosen to attain a pre-specified FPR level, say $u_0$, we may obtain STEAM estimators of $\PPV_0(\cbar\subut)$ and $\NPV_0(\cbar\subut)$ 
as $\PPVhat\supSTEAM_0(\chat\subut)$ and $\NPVhat\supSTEAM_0(\chat\subut)$, where $\chat\subut = {\FPRhat\supSTEAMinv}_0(u_0)$
\begin{align*}
        \PPVhat\supSTEAM_0(c) &= \frac{\muhat_0\ \TPRhat\supSTEAM_0(c) }{\muhat_0\ \TPRhat\supSTEAM_0(c) + (1-\muhat_0)\ \FPRhat\supSTEAM_0(c)} \\
        \NPVhat\supSTEAM_0(c) &= \frac{(1-\muhat_0)\ \{1-\FPRhat\supSTEAM_0(c) \}}{(1-\muhat_0)\ \{1-\FPRhat\supSTEAM_0(c) \}+\muhat_0\ \{1-\TPRhat\supSTEAM_0(c) \}}.
\end{align*}
$\muhat_0=N_t^{-1}\sum_{i\in \IT}\mthatbetahat(\Pthatbetahati)$ is the SS estimator of the prevalence $\mu_0=\Pbb_0(Y=1)$. 

Similar to the supervised setting, the above STEAM estimator may suffer from overfitting bias since the labeled data are used for both estimating the outcome model and the calibrated conditional risk function. Following similar strategies given in \cite{gronsbell_semi_2018}, it is not difficult to employ a $K$-fold cross-validation (CV)  procedure to correct for the overfitting bias in the STEAM estimators. Detailed forms of the CV estimator can be found in the Appendix C. For all numerical studies, we employ the CV correction for the STEAM estimator. 

\subsection{Asymptotic Results for STEAM Estimators}
Though analogous results hold for all STEAM estimators, we present the main results for $\TPRhat_0\supSTEAM$.
\begin{thm}
Under assumptions outlined in the Appendix, when either \eqref{eq:mu} or \eqref{eq:pi} is correct, the $\TPRhat_0\supSTEAM(c)$ is consistent for $\TPR_0(c)$. Moreover, 
$$
\nhalf\left\{\TPRhat_0\supSTEAM (c) - \TPR_0 (c) \right\} \rightsquigarrow \mathcal N(0,  \mathcal V)
$$
\end{thm}
The proof for consistency directly follows from Remark 1 and is outlined in Appendix A. To show asymptotic normality of STEAM estimators, let
$$
\TPRhat\supSTEAM_0(c) = \frac{N_t^{-1}\sum_{i\in\mathcal I_T}  I(\Pthatbetahati \ge c) \mthat(\Pthatbetahati,\Pthatbetahat,\what(\cdot, \balphahat,\bbetahat))}{N_t^{-1}\sum_{i\in\mathcal I_T}  \mthat(\Pthatbetahati,\hat{\mathcal P}_{\bbetahat},\what(\cdot, \balphahat,\bbetahat))}  =\frac{\widehat{\xi}(c, \bbetahat)}{\widehat{\xi}(0, \bbetahat)} ,
$$
$\xi\supalpha(c,\bbeta)=E\{I(\Ptbetai \ge c) m_0(\Ptbetai,\Ptbeta,w(\cdot, \balpha, \bbeta) \}$, and $\xi(c, \bbeta) = \xi\supalphabar(c,\bbeta)$. We show that
$$\nhalf\{\xihat(c,\bbetahat)-\xi(c,\bbetabar)\}=\nnhalf \sum_{j\in\mathcal I_L} \left[ \wbar_j I(\Ptbetabarj > c)  \{Y_j  -\xi(c,\bbetabar )\} +  \bC_2\trans\bphi_i\supbeta \right]+ o_p(1)$$
where $\bphi_i\supbeta$ is the influence function of $\bbetahat$, and $\bC_2$ is defined in Appendix B. The weak convergence of $\nhalf\{\TPRhat\supSTEAM_0(c)-\TPR_0(c)\}$ thus follows similar as Appendix A of \textcite{gronsbell_semi_2018}.

 Even though the asymptotic variance can be obtained through influence function, a direct estimate is difficult because it involves unknown conditional density functions which are difficult to estimate explicitly, particularly under model mis-specification. We instead propose the use of a simple resampling procedure for inference based on \textcite{jin_simple_2001} in the next section.

\subsection{Perturbation Resampling Procedure for Inference}
Now we outline the perturbation resampling procedure for inference. Let $\mathcal G = \{G_i: i=1, \dots, N \}$ be non-negative independent and identically distributed random variables with unit mean and variance that are independent of the observed data. We first obtain perturbed estimates $\bbetahat^\ast$  as:
$$
\bbetahat^\ast = \argmax_{\bm\beta} \left\{n^{-1} \sum_{i\in\IL} \ell_g(\bbeta;Y_i,\bX_i)G_i - \lambda_{\mu,n}  \|(\bbeta_{-1}/\bbetatilde^\ast_{-1})^\gamma\|_1 \right\}
$$
where $\bbetatilde^\ast$ is perturbed initial estimates obtained from analogously perturbing its estimating equations. This leads to the perturbed estimates of the density ratio weights: 
$$
\what_i^* \equiv 
\what(\bX_i; \balphahat,\bbetahat^*) = \frac{1-\pihat(\bX_i; \balphahat, \bbetahat^*)}{\pihat(\bX_i; \balphahat, \bbetahat^*)}
$$
where
$$
\pihat(\bX; \balphahat, \bbetahat^\ast) = \frac{\sum_{i\in \IU \cup \IT} K_{\pi,h_1}\left\{(\balphahat, \bbetahat^\ast)\trans  (\bX_i-\bX)\right\}S_i}{\sum_{i\in \IU \cup \IT}  K_{\pi,h_1}\left\{(\balphahat, \bbetahat^\ast)\trans  (\bX_i-\bX)\right\}} 
$$
Since $\balphahat$ and $\pihat(\cdot)$ are estimated based $\mathcal U \cup \mathcal T$, their contribution to the asymptotic variance is of higher order when $N\gg n$.
We then obtain the perturbed estimate of conditional risk function as
$$
\mthatbetahatstar(q) = \frac{\sum_{i\in\IL}K_{m,h_2}\{ \Pthatbetahatstar(\bX_i)-q\}\what^\ast_i Y_i G_i}{\sum_{i\in\IL}K_{m.h_2}\{\Pthatbetahatstar(\bX_i)-q\}\what^\ast_i G_i} 
$$
Finally we calculate the perturbed $\TPRhat_0\supSTEAM$ as
$$
\TPRhat\supSTEAMstar_0(c) = \frac{\sum_{i\in\IT}  I(\Pthatbetahatstari \ge c) \mthatbetahatstar(\Pthatbetahatstari)}{\sum_{i\in\IT}  \mthatbetahatstar(\Pthatbetahatstari)}  
$$
It can be shown using arguments from \textcite{jin_simple_2001} that the asymptotic distribution of $\nhalf\left\{\TPRhat\supSTEAM_0(c)-\TPR_0(c)\right\}$ coincides with that of $\nhalf\left\{\TPRhat\supSTEAMstar_0(c)-\TPRhat\supSTEAM_0(c)\right\}|\mathcal D$. This allows us to consistently approximate the SE of $\TPRhat\supSTEAM(c)$ based on the empirical standard deviation of resamples $\TPRhat\supSTEAMstar(c)$. The confidence intervals of $\TPRhat\supSTEAM$ an be constructed by using the empirical percentiles of the resamples. We can obtain perturbed counterparts of $\FPRhat\supSTEAM(c)$, $\ROChat\supSTEAM(c)$, $\PPVhat\supSTEAM(c)$, $\NPVhat\supSTEAM(c)$ respectively denoted as $\FPRhat\supSTEAMstar(c)$, $\ROChat\supSTEAMstar(c)$, $\PPVhat\supSTEAMstar(c)$, $\NPVhat\supSTEAMstar(c)$, and obtain SE estimates and construct confidence intervals accordingly. 

For practical use, an issue of this perturbation resampling procedure is that calculating $\pihat(\bX; \balphahat, \bbetahat^\ast)$ requires two-dimensional smoothing, which can be very time-consuming. To overcome this difficulty, we propose an approximated resampling procedure that is more computationally feasible. We first generate $\{\bm{\mathcal G}^{[b]}: (G_1^{[b]}, \cdots, G_N^{[b]})\trans , b=1,\cdots, B\}$ where $\bm{\mathcal G}$ has dimension $N\times B$, followed by the perturbed $\bbeta$ estimates $\{\bbetahat^*: (\bbetahat_1^{*[b]}, \cdots, \bbetahat_p^{*[b]}), b=1,\cdots,B\}$ where $\bbetahat^*$ has dimension $p\times B$. We then approximate the perturbed estimates $\pihat$ by
$$
[\pihat(\bX_i; \balphahat, \bbetahat^*)]_{N\times B} = [\pihat(\bX_i; \balphahat, \bbetahat)]_{N\times B}+ [\nabla \pihat(\bX_i; \balphahat, \bbetahat)]_{N\times p} [\bbetahat^* - \bbetahat]_{p \times B},
$$
where
$$
\nabla \pihat(\bX; \balphahat, \bbeta) = \frac{\partial \pihat(\bX; \balphahat, \bbeta)}{\partial \bbeta}.
$$
The perturbed estimate of conditional risk can then be obtained as $[\mthatbetahatstar(\Pthatbetahatstari)]_{N\times B}$, and the perturbed $\TPRhat\supSTEAMstar_0(c)$ thus follows. We evaluate the performance of the perturbation resampling procedure with and without approximation through simulations.

\section{Simulation Studies}
We performed extensive simulations to assess the finite sample bias, standard error (SE), root mean square error (RMSE) of our proposed estimators (STEAM) compared to alternative estimators. In separate simulations we also examined the performance of the perturbation procedure, with and without approximation, for inference based on STEAM estimators. Across all numerical studies including the data application in section 5, we used the adaptive LASSO to estimate $\balpha$ and $\bbeta$, where we choose $\lambda_{\mu,{n}}$ and  $\lambda_{\pi,N}$ using a modified BIC criteria (\cite{minnier_perturbation_2011}). For the bandwidth in the two-dimensional smoothing for $\pi(\cdot)$, we use a Gaussian product kernel of order with a plug-in bandwidth $h_1 = \widehat\sigma N^{-1/6}$, where $\widehat\sigma$ is the sample standard deviation of either $\balphahat\trans \bX_i$, $\bbetahat\trans  \bX_i$. Prior to smoothing, $\balphahat\trans \bX_i$ and $\bbetahat\trans \bX_i$ were standardized and transformed by a probability integral transform based on the normal cumulative distribution function to induce approximately uniformly distributed inputs, which can improve finite-sample performance (\cite{wand_transformations_1991}). For the non-parametric smoothing for $m(\cdot)$, we used the Gaussian kernel with $h_2=n^{-0.4}\widehat\sigma_{\mathcal P}$, where $\widehat\sigma_{\mathcal P}$ is the empirical standard deviation of $\{\widehat{\mathcal P_i}\}_{i=1}^n$. As we focused on binary outcomes, we specified logistic link functions $g_\mu(u) = g_\pi (u) = 1/(1+e^{-u})$.

For comparison, we considered alternative estimators for performance accuracy parameters: (1) the naive estimators, where the performance accuracy parameters are estimated using cross-validation on the labeled source dataset (\textit{source}), (2) estimation based on $100$ randomly labeled validation samples in target dataset (\textit{target\_labeled}), (3) the standard weighted estimators based on cross-validation (\textit{weighted}), (4) the doubly robust estimator proposed by \textcite{alonzo_assessing_2005} with cross-validation (\textit{DR-aug}). This estimator is a special case of \textcite{rotnitzky_doubly_2006} and \textcite{fluss_estimation_2009} when $Y$ is independent of $S$ conditional on $\bX$. 

Throughout, we generated $\bX$ from $MVN\{\sigma^2(1-\rho)\mathbb I_{p\times p}+\sigma^2\rho \}$, $S=1|\bX \sim Bernoulli\{\pi(\bX) \}$, and $Y=1|\bX \sim Bernoulli\{\mu(\bX) \}$. We considered simulating data that roughly resembled the EHR data example in terms of model parameters but were simple enough to be more broadly relevant. To this end, we considered $p=10$ covariates with variance $\sigma^2=1$, and correlations $\rho=0.2$. These simulations were varied over different model specifications, predictive strength of $\bX$ to $S$, and sample sizes. We considered the following data generating mechanisms:
\begin{align*}
    \mu(\bX) &= expit (-0.25+0.8X_1+0.8X_2+0.4X_3+0.4X_4+0.2X_1X_2-0.1X_2X_3+0.2X_3X_4)\\
    \pi(\bX) &= \begin{cases}
    expit (0.1X_1+0.05X_2-0.1X_5-0.05X_6+0.05X_1X_2) & \text{(Weak)}\\ expit (0.2X_1+0.1X_2-0.2X_5-0.1X_6+0.1X_1X_2) &\text{(Moderate)} \\ expit (0.6X_1+0.3X_2-0.6X_5-0.3X_6+0.3X_1X_2) &\text{(Strong)}
    \end{cases}
\end{align*}
Either $\pi(\bX;\balpha)$ or $\mu(\bX;\bbeta)$ were potentially mis-specified by neglecting the pairwise interaction terms in model fitting:
\begin{itemize}
    \item[(1)] \textit{Both correct:} Fitting $\mu(\bX)=expit(\bbeta\trans \bm Z_\mu)$ and $\pi(\bX)=expit(\balpha\trans \bm Z_\pi)$.
    \item[(2)] \textit{$\mu(\bX)$ correct and $\pi(\bX)$ mis-specified:} Fitting $\mu(\bX)=expit(\bbeta\trans \bm Z_\mu)$ and $\pi(\bX)=expit(\balpha\trans \bX)$.
    \item[(3)] \textit{ $\pi(\bX)$ correct and $\mu(\bX)$ mis-specified:} Fitting $\mu(\bX)=expit(\bbeta\trans \bX)$ and $\pi(\bX)=expit(\balpha\trans \bm Z_\pi)$.
\end{itemize}
where $\bm Z_\mu = (\bm 1, \bX, X_1X_2, X_2X_3, X_3X_4)\trans $, $\bm Z_\pi = (\bX, X_1X_2)\trans $. The outcome prevalence was approximately $45\%$, and the sample size in source and target are roughly equal. We considered $N=10000$ and $n=200, 400$. The results in each scenario are summarized from $1000$ simulated datasets.

We present results for $\ROC(u_0)$, $\PPV(c_{u_0})$, $\NPV(c_{u_0})$ for $u_0=0.05$. Table \ref{table:RMSE} presents the bias, SE, and RMSE across mis-specification scenarios with moderate strength of $\bX$ to $S$ with $N=10000$. As expected, the \textit{source} estimators exhibit noticeable bias, regardless of the size of $n$. This is consistent with the consensus that the naive estimator based on source data is not appropriate for prediction performance assessment in the presence of covariate shift. The \textit{target\_labeled} estimators are unbiased, but have large SEs dues to the limited sample size. The \textit{weighted} estimators are single robust and have noticeable bias when $\pi(\bX)$ is mis-specified. Among the three model mis-specification scenarios considered, the bias of STEAM is small relative to the RMSE, verifying its doubly-robustness. STEAM achieves the lowest RMSE uniformly.

\begin{figure}
\centering
\includegraphics[width=1.1\textwidth]{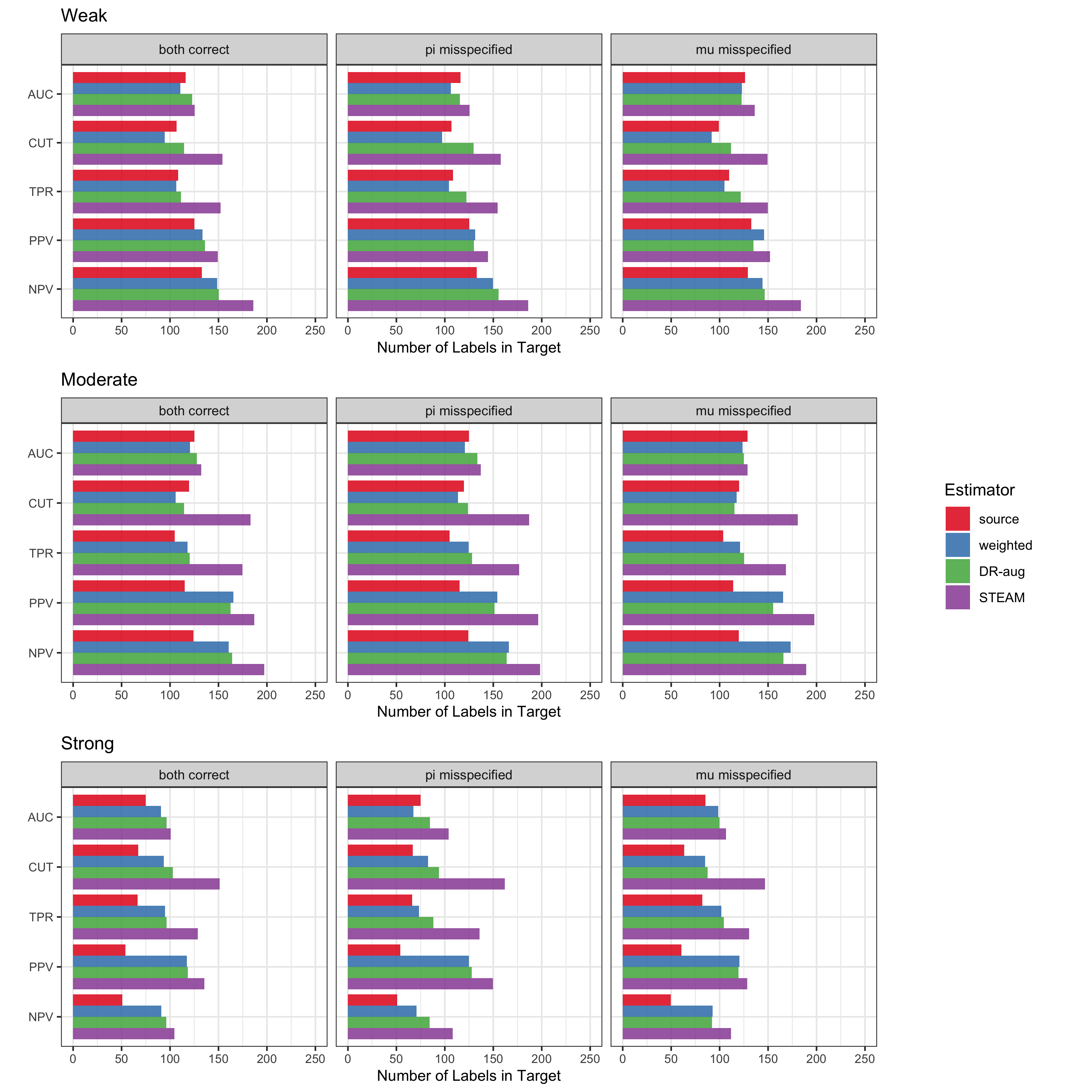}
\caption{Number of labels in target cohort required to achieve the same root mean square errors (RMSE) as each estimator, by three model mis-specification scenarios for different predictive strength of $\bX$ to $S$ (Weak, Moderate, Strong), over 1000 simulated datasets. Larger number of labels denotes greater efficiency (lower MSE) relative to \textit{target\_labeled}. $N=10000$, $n=200$, $k=5$.}
\label{fig:RE}
\end{figure}

Figure \ref{fig:RE} presents the number of labels in target cohort needed to achieve the same root mean square error (RMSE) as each estimator across different predictive strength of $\bX$ to $S$ (Weak, Moderate, Strong), and mis-specification scenarios for $N=10000$, $n=200$, $k=5$. A larger number of labels needed in target cohort represents lower RMSE and higher efficiency of the estimator. The STEAM estimators are more efficient than \textit{target\_labeled}, \textit{weighted} and \textit{DR-aug} across the settings. The amount of efficiency gain of STEAM over creating labels in target varies with the predictive strength of $\bX$ to $S$. We observe that when $\bX$ is weakly or moderately correlated with $S$, using the $n=200$ labels in source for STEAM estimators is equivalent to creating $\sim120-200$ new labels in target in terms of RMSE, but when $\bX$ is strongly associated with $S$, using STEAM is less advantageous, only equivalent to $\sim100-150$ labels in target in terms of RMSE. 

To implement the perturbation procedure, we used the weights $G_i\sim 4\times Beta(0.5, 1.5)$ and $1000$ sets of $\mathcal G$ for SE and CI estimation. We considered evaluating the perturbations only in the scenario where both $\pi(\bX)$ and $\mu(\bX)$ are correctly specified and $\bX$ has moderate predictive strength to $S$, with $N=10000$ and $n=200$. The results are presented in Table \ref{table:SE}. The SEs estimated with and without approximation in perturbation approximated well the empirical standard error. The coverage of the $95\%$ CIs were also close to nominal levels.

\section{Application to Rheumatoid Arthritis (RA) Phenotyping model}
To illustrate the performance of our proposed estimators, we applied our procedures to evaluate a phenotyping algorithm for classifying RA disease, using EHR data from Partner's Healthcare (\cite{liao_electronic_2010}). The source dataset from 2009 consists of 20,451 total patients, of which $n=267$ were labeled with the true disease status through chart review by an rheumatologist. Both narrative and codified data were available to develop the prediction model ($p = 21$). The narrative variables, including disease diagnoses and medications, were obtained with natural language processing. The codified data included ICD-9 codes, electronic prescriptions, and laboratory values. Healthcare utility was also included. The transformation $u \mapsto \log(1+u)$ was applied to all count variables to mitigate instability in the estimation due to skewness in their distributions. An RA phenotyping model, named as \textit{ALASSO-2009}, was trained on labeled data from 2009. We are interested in assessing its performance in classifying the disease status of patients in the same EHR database across different time windows, in 2011 ($N_t=25,405, n_t=100$), 2013 ($N_t=30,804, n_t=100$), 2015 ($N_t=36,095, n_t=150$) and 2017 ($N_t=39,550, n_t=200$). Each of these consists of a large number of unlabeled data with the same set of covariates available. A random subset of each dataset was labeled with true RA status, and can serve as the validation sets to verify our results.

\begin{figure}
\centering
\includegraphics[width=0.9\textwidth]{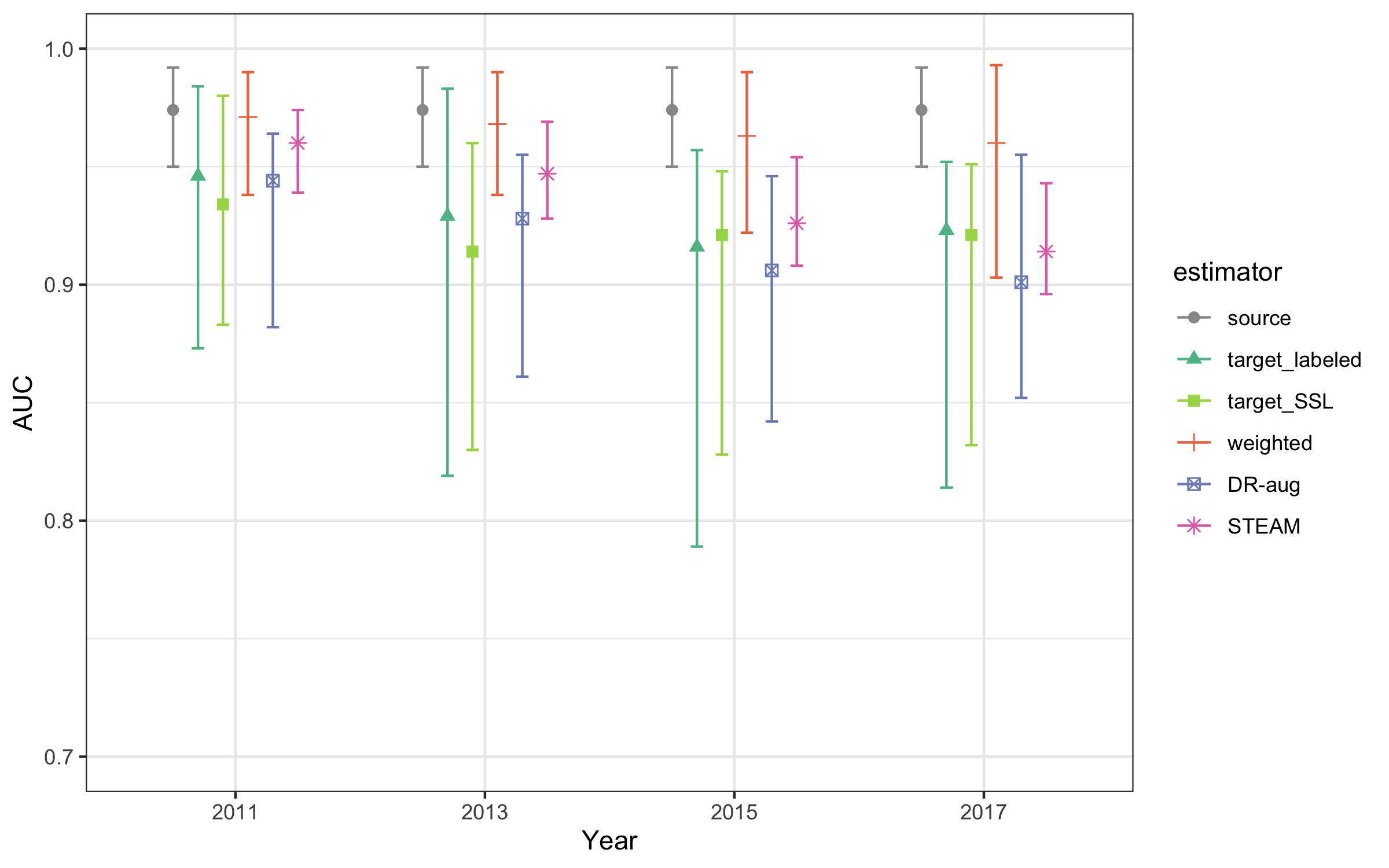}
\caption{Estimated AUC with $95\%$ CI of \textit{ALASSO-2009} on datasets in 2011, 2013, 2015 and 2017 based on different methods.}
\label{fig:RA_AUC}
\end{figure}

For comparison, we considered \textit{source}, \textit{target\_labeled}, \textit{weighted} and \textit{DR-aug} estimators as in Section 4. We also 
considered semi-supervised estimation using labeled and unlabeled data in target as proposed by \textcite{gronsbell_semi_2018} (\textit{target\_SSL}). The estimated AUC of the model on the target datasets are presented in Figure \ref{fig:RA_AUC}. The AUC estimates for the target datasets obtained using the validation sets (\textit{target\_labeled} and \textit{target\_SSL}) are noticeably different from the ones estimated from the source data (\textit{source}), especially in 2015 and 2017. This is not too surprising since many changes in the database have occurred over the years: for instance, patient characteristics, such as age, have undergo changes. Moreover, the EHR system was switched to Epic (Epic System Corporation) and the International Classification of Diseases (ICD) system was changed from version 9 to version 10 around 2015-2016, which may also explain the lower AUC in 2015 and 2017. The  point estimates based on our proposed STEAM method are similar to \textit{target\_labeled}, suggesting the stability of the proposed procedure in a real data setting. Moreover, we observe substantial efficiency gains of STEAM over \textit{target\_labeled}, \textit{weighted} and \textit{DR-aug}. As a result, we have a more precise estimate of the prediction performance of the phenotype algorithm using the STEAM method. The estimated ROC curves along with 95\% point-wise confidence bands can be found in Appendix D. 

The estimates of cutoff values and the corresponding $TPR$, $PPV$ and $NPV$ estimates of the binary classification rule for $u_0 = 0.05$ and $0.10$ as well as their $95\%$ CIs are presented in Figure \ref{fig:RA_others}. For choosing the cutoff values, STEAM is $\sim 4.2-7.2$ times as efficient as \textit{target\_labeled}, and $\sim 1.1-2.1$ times as efficient as \textit{DR-aug}. For each accuracy measure, STEAM estimators are $3.1-5.8$ times as efficient as \textit{target\_labeled}, and the efficiency gain is at least $14\%$ with gains as high as $245\%$ compared with \textit{DR-aug}.

\section{Discussion}
In this paper, we proposed a robust and efficient transfer learning procedure for model evaluation rather than model fitting. Our estimators are doubly robust, i.e. they are consistent when either the prediction model $\mu(\bX)$ or the weighting model $\pi(\bX)$ is correct.  We addressed potential overfitting bias in our SS estimators with CV and also developed a perturbation resampling procedure for making inference. We illustrate the performance and practical utility of the proposed estimators through simulations and application on an RA phenotying model. 

We have assumed that the true outcomes are labeled completely at random in source population, which may be reasonable if investigators control the labeling. However, this assumption could be restrictive if labeling was stratified by some known factors or if some records that are available were not labeled for research purposes. Further work is needed to extend our results to the missing at random (MAR) setting (\cite{rubin_inference_1976}) to allow the labeling process to depend on $\bX$. Other refinements and extensions to the proposed approach is possible. For instance, when the prediction model is clearly mis-specified, we may incorporate an additional outcome model in the estimation procedure to preserve the doubly robust property, i.e. the estimator are consistent if either the new outcome model or the weighting model is correct. 

Throughout, we focused on the setting with fixed $p$ but accommodated settings in which $p$ is not small relative to $n$ in finite sample with regularized estimation. However, when $p$ is large so that $\nu$ is large, a large power parameter $\gamma$ would be required to maintain the oracle properties, leading to an unstable penalty and poor finite sample performance. Estimation under the setting with $p \gg n$ would require different theoretical justifications and warrant additional research.

\clearpage


\printbibliography
\clearpage
\newgeometry{left=1.5cm, right=1.5cm, top=1cm, bottom=2cm}
\begin{table}
    \centering
    \scriptsize
    \begin{tabular}{l l l c c c c c c c c c}
\toprule
& & & \multicolumn{3}{c}{both correct}&
\multicolumn{3}{c}{$\pi(X)$ mis-specified} &
\multicolumn{3}{c}{$\mu(X)$ mis-specified}\\
\cmidrule(r){4-6}\cmidrule(l){7-9}\cmidrule(l){10-12}
& & Estimator & Bias & SE & RMSE & Bias & SE & RMSE & Bias & SE & RMSE\\
\bottomrule
Cutoff &  $n=200$ & source & 1.02 & 4.24 & 4.35 & 1.02 & 4.24 & 4.35 & 1.24 & 4.19 & 4.36 \\
& & target\_labeled & -0.19 & 4.86 & 4.86 & -0.19 & 4.86 & 4.86 & -0.40 & 4.88 & 4.89 \\
& & weighted & -0.33 & 4.64 & 4.64 & -0.34 & 4.47 & 4.47 & -0.05 & 4.42 & 4.42  \\
& & DR-aug & -0.53 & 4.42 & 4.45 & -0.55 & 4.23 & 4.27 & -0.35 & 4.45 & 4.46  \\
& & STEAM & -0.21 & 3.47 & 3.47 & -0.39 & 3.42 & 3.43 & -0.06 & 3.52 & 3.51 \\
\cmidrule(r){2-12}
& $n=400$ & source & 1.94 & 3.04 & 3.59 & 1.94 & 3.04 & 3.59 & 1.80 & 3.17 & 3.64 \\
& & target\_labeled & -0.40 & 5.40 & 5.40 & -0.40 & 5.40 & 5.40 & -0.34 & 5.25 & 5.25 \\
& & weighted & 0.18 & 3.05 & 3.05 & 0.14 & 3.06 & 3.06 & 0.04 & 3.12 & 3.12  \\
& & DR-aug & 0.22 & 2.91 & 2.92 & 0.35 & 2.91 & 2.93 & 0.25 & 3.11 & 3.12 \\
& & STEAM &  0.08 & 2.52 & 2.52 & 0.01 & 2.49 & 2.48 & 0.09 & 2.53 & 2.53 \\
\bottomrule
AUC &  $n=200$ & source & 1.16 & 3.66 & 3.83 & 1.16 & 3.66 & 3.83 & 0.90 & 3.68 & 3.78 \\
& & target\_labeled & 0.57 & 4.74 & 4.76 & 0.57 & 4.74 & 4.76 & 0.62 & 4.60 & 4.63 \\
& & weighted & 0.39 & 3.89 & 3.90 & 0.97 & 3.79 & 3.90 & 0.17 & 3.87 & 3.86  \\
& & DR-aug & 0.41 & 3.73 & 3.75 & 0.52 &  3.66 & 3.70 & 0.72 & 3.77 & 3.84 \\
& & STEAM & -0.38 & 3.71 & 3.72 & -0.10 & 3.65 & 3.65 & -0.68 & 3.73 & 3.78 \\
\cmidrule(r){2-12}
& $n=400$ & source & 1.19 & 2.49 & 2.76 & 1.19 & 2.49 & 2.76 & 1.10 & 2.53 & 2.75 \\
& & target\_labeled & 0.59 & 4.70 & 4.73 & 0.59 & 4.70 & 4.73 & 0.54 & 4.70 & 4.72 \\
& & weighted &  0.37 & 2.63 & 2.65 & 1.06 & 2.64 & 2.84 & 0.26 & 2.66 & 2.66 \\
& & DR-aug & 0.40 & 2.58 & 2.61 & 0.45 & 2.61 & 2.64 & 0.33 & 2.61 & 2.63 \\
& & STEAM & -0.36 & 2.51 & 2.52 & -0.18 & 2.52 & 2.52 & -0.27 & 2.53 & 2.53 \\
\bottomrule
TPR &  $n=200$ & source & 3.60 & 9.20 & 9.88 & 3.60 & 9.20 & 9.88 & 3.33 & 9.36 & 9.93 \\
& & target\_labeled & 0.43 & 10.52 & 10.53 & 0.43 & 10.52 & 10.53 & 0.59 & 10.57 & 10.59 \\
& & weighted & 0.82 & 9.26 & 9.3 & 1.87 & 8.84 & 9.04 & 0.73 & 9.14 & 9.17  \\
& & DR-aug & 0.80 & 9.18 & 9.21 & 0.78 & 8.88 & 8.91 & 0.25 & 9.02 & 9.02 \\
& & STEAM & 0.22 & 7.60 & 7.60 & 0.66 & 7.52 & 7.55 & -0.12 & 7.75 & 7.75 \\
\cmidrule(r){2-12}
& $n=400$ & source & 2.63 & 6.52 & 7.03 & 2.63 & 6.52 & 7.03 & 2.84 & 6.89 & 7.44 \\
& & target\_labeled & 0.39 & 11.35 & 11.36 & 0.39 & 11.35 & 11.42 & 0.35 & 11.21 & 11.22 \\
& & weighted &  0.58 & 6.47 & 6.5 & 1.83 & 6.49 & 6.88 & 1.17 & 6.53 & 6.62  \\
& & DR-aug & 0.50 & 6.14 & 6.16 & 0.73 & 6.11 & 6.15 & 0.36 & 6.44 & 6.45 \\
& & STEAM & -0.32 & 5.62 & 5.63 & -0.15 & 5.59 & 5.58 & -0.36 & 5.69 & 5.68 \\
\bottomrule
PPV &  $n=200$ & source & 3.66 & 3.82 & 5.29 & 3.66 & 3.82 & 5.29 & 3.66 & 3.88 & 5.33 \\
& & target\_labeled & 0.21 & 5.79 & 5.78 & 0.21 & 5.78 & 5.78 & 0.82 & 5.37 & 5.43 \\
& & weighted & 0.89 & 4.24 & 4.33 & 2.04 & 4.01 & 4.50 & 1.24 & 4.16 & 4.34  \\
& & DR-aug & -0.79 & 4.30 & 4.37 & -0.81 & 4.49 & 4.55 & -0.32 & 4.49 & 4.50 \\
& & STEAM & -0.41 & 4.02 & 4.04 & -0.01 & 3.93 & 3.93 & -0.57 & 3.89 & 3.93 \\
\cmidrule(r){2-12}
& $n=400$ & source & 2.99 & 2.24 & 3.74 & 2.99 & 2.24 & 3.73 & 3.06 & 2.36 & 3.87 \\
& & target\_labeled & 0.52 & 5.37 & 5.40 & 0.52 & 5.37 & 5.38 & 0.35 & 5.50 & 5.49 \\
& & weighted & 0.83 & 2.53 & 2.66 & 1.46 & 2.54 & 2.93 & 1.10 & 2.53 & 2.75  \\
& & DR-aug & -0.42 & 2.64 & 2.67 & -0.34 & 2.88 & 2.90 &-0.41 & 2.69 & 2.72 \\
& & STEAM & -0.31 & 2.44 & 2.45 & -0.25 & 2.45 & 2.46 & -0.29 & 2.46 & 2.47 \\
\bottomrule
NPV &  $n=200$ & source & -2.26 & 4.63 & 5.14 & -2.26 & 4.63 & 5.14 & -2.35 & 4.65 & 5.20 \\
& & target\_labeled & 0.56 & 5.87 & 5.88 & 0.56 & 5.87 & 5.88 & 0.76 & 5.88 & 5.92 \\
& & weighted & -0.28 & 4.50 & 4.50 & 0.17 & 4.43 & 4.42 & -0.48 & 4.28 & 4.30  \\
& & DR-aug & -0.42 & 4.43 & 4.45 & -0.25 & 4.44 & 4.45 & -0.38 & 4.38 & 4.40 \\
& & STEAM & -0.30 & 4.04 & 4.04 & -0.13 & 4.04 & 4.03 & -0.39 & 4.10 & 4.11 \\
\cmidrule(r){2-12}
& $n=400$ & source & -2.47 & 3.70 & 4.44 & -2.47 & 3.70 & 4.44 & -2.39 & 3.71 & 4.41 \\
& & target\_labeled & 0.77 & 5.86 & 5.89 & 0.77 & 5.86 & 5.89 & 0.69 & 5.91 & 5.93 \\
& & weighted & -0.32 & 3.40 & 3.40 & 0.19 & 3.44 & 3.43 & -0.26 & 3.34 & 3.34  \\
& & DR-aug & -0.51 & 3.27 & 3.31 & -0.30 & 3.37 & 3.38 & -0.40 & 3.37 & 3.39 \\
& & STEAM & -0.33 & 3.17 & 3.17 & -0.24 & 3.17 & 3.17 & -0.35 & 3.16 & 3.17 \\
\bottomrule
\end{tabular}
    \caption{Bias, SE, and RMSE of estimators under different model mis-specification scenarios with moderate association of $\bX$ to $S$, over 1000 simulated datasets. $N=10000$, $k=5$, $n=200$ or $400$. All values are multiplied by 100.  }
    \label{table:RMSE}
\end{table}
\clearpage
\restoregeometry

\begin{table}
    \centering
    \begin{tabular}{l c c c c c c}
\toprule
& Estimate & ESE & ASE & CovProb & ASE\_approx & CovProb\_approx \\
\bottomrule
Cutoff & 0.678 & 0.040 & 0.037 & 0.938 & 0.037 & 0.925 \\
AUC & 0.779 & 0.032 & 0.030 & 0.929 & 0.036 & 0.968 \\
TPR & 0.286 & 0.053 & 0.051 & 0.920 & 0.055 & 0.929 \\
PPV & 0.823 & 0.032 & 0.033 & 0.944 & 0.036 & 0.960 \\
NPV & 0.620 & 0.040 & 0.036 & 0.918 & 0.043 & 0.964 \\
\bottomrule
\end{tabular}
    \caption{Performance of perturbation resampling for STEAM in 1000 simulated datasets when both $\pi(\bX; \balpha)$, $\mu(\bX; \bbeta)$ are correctly specified. ASE: average of estimated SE based on the standard deviation of perturbed estimates; CovProb: coverage probability of $95\%$ CIs; ASE: average of estimated SE based on the standard deviation of perturbed estimates using the approximated method; CovProb: coverage probability of $95\%$ CIs using the approximated method. $N=10000$, $n=200$, $k=5$.}
    \label{table:SE}
\end{table}
\clearpage

\clearpage

\begin{figure}
\centering
\includegraphics[width=\textwidth]{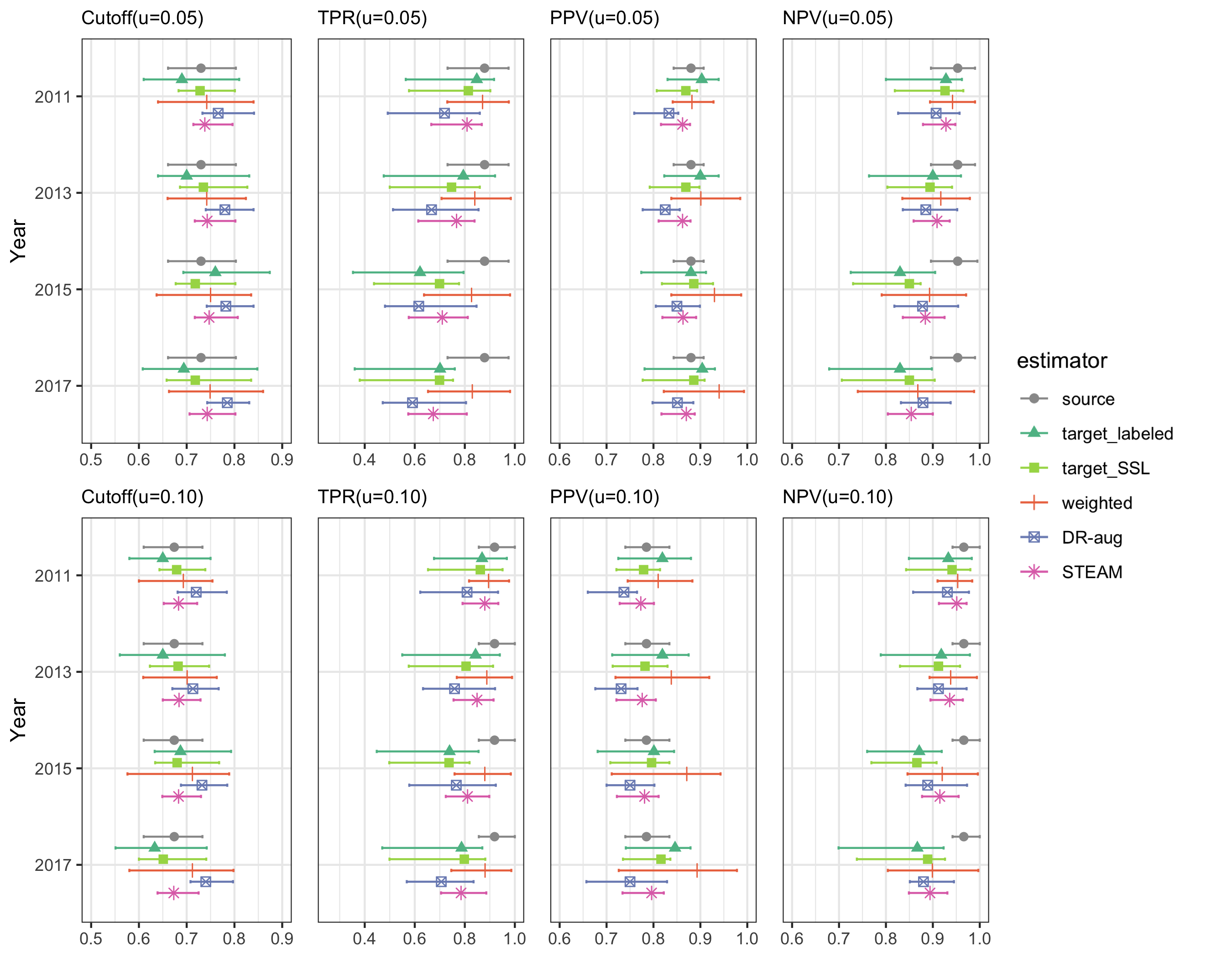}
\caption{Estimated cutoffs and the corresponding $TPR$, $PPV$ and $NPV$ estimates of the binary classification rule for $u_0=0.05$ (top) and $u_0=0.10$ (bottom) of \textit{ALASSO-2009}. Error bars denote $95\%$ CIs.}
\label{fig:RA_others}
\end{figure}

\clearpage

\def\muhat{\widehat{\mu}}
\def\PPVhat{\widehat{\PPV}}
\def\NPVhat{\widehat{\NPV}}
\def\ROChat{\widehat{\ROC}}
\def\AUChat{\widehat{\AUC}}
\def\TPRhat{\widehat{\TPR}}
\def\FPRhat{\widehat{\FPR}}
\def\supwgt{^{\scriptscriptstyle \sf wgt}}
\def\TPR{\rm{TPR}}
\def\FPR{\rm{FPR}}
\def\ROC{\rm{ROC}}
\def\PPV{\rm{PPV}}
\def\NPV{\rm{NPV}}
\def\AUC{\rm{AUC}}
\def\bbeta{\bm\beta}
\def\balpabar{\bm{\bar\alpha}}
\def\balphatilde{\bm{\widetilde\alpha}}
\def\bbetatilde{\bm{\widetilde\beta}}
\def\betatilde{\widetilde\beta}
\def\bbetahat{\bm{\widehat\beta}}
\def\bbetabar{\bm{\bar\beta}}
\def\balpha{\bm\alpha}
\def\balphahat{\bm{\widehat\alpha}}
\def\balphabar{\bm{\bar\alpha}}
\def\bX{\bm X}
\def\bx{\bm x}
\def\IL{\mathcal I_L}
\def\IU{\mathcal I_U}
\def\IT{\mathcal I_T}
\def\pto{\overset{P}\to}
\def\half{\frac{1}{2}}
\def\ninv{n^{-1}}
\def\nhalf{n^{\half}}
\def\nnhalf{n^{-\half}}
\def\Nhalf{N^{\half}}
\def\Nnhalf{N^{-\half}}
\def\Nthalf{N_t^{\half}}
\def\Ntinv{N_t^{-1}}
\def\Ntnhalf{N_t^{-\half}}
\def\Nhalf{N^{\half}}
\def\Isc{\mathcal{I}}
\def\Fscdot{\dot{\Fsc}}
\def\Fsc{\mathcal{F}}
\def\Fschat{\widehat{\Fsc}}
\def\xihat{\widehat{\xi}}
\def\Psc{\mathcal{P}}
\def\Tsc{\mathcal{T}}
\def\Ptbetabar{ {\mathcal P}_{0}\supbetabar}
\def\Ptbetabari{ {\mathcal P}_{0,i}\supbetabar}
\def\Ptbetabarj{ {\mathcal P}_{0,j}\supbetabar}
\def\Ptbetahat{ {\mathcal P}_{0}\supbetahat}
\def\Ptbetahati{ {\mathcal P}_{0,i}\supbetahat}
\def\Pthatbetahat{\widehat{\Psc}_{0}\supbetahat}
\def\Pthatbetahatstar{\widehat{\Psc}_{0}\supbetahatstar}
\def\Pthatbetahati{\widehat{\Psc}_{0,i}\supbetahat}
\def\Pthatbetahatstari{\widehat{\Psc}_{0,i}\supbetahatstar}
\def\Pthatbetahatnki{\widehat{\Psc}_{0,i}\supbetahatnk}
\def\Pthatbetabar{\widehat{\Psc}_{0}\supbetabar}
\def\Pthatbetabari{\widehat{\Psc}_{0,i}\supbetabar}
\def\Pthatbetai{\widehat{\Psc}_{0,i}\supbeta}
\def\Pthatbetaj{\widehat{\Psc}_{0,j}\supbeta}
\def\Ptbeta{\Psc_{0}\supbeta}
\def\Ptbetai{\Psc_{0,i}\supbeta}
\def\Ptbetaj{\Psc_{0,j}\supbeta}
\def\mtbetabar{m_{0}\supbetabar}
\def\mthatbetahat{\mthat\supbetahat}
\def\mthatbetabar{\mthat\supbetabar}
\def\mthatbetahatstar{\mthat\supbetahatstar}
\def\Mthat{\widehat{M}_0}
\def\mthat{\widehat{m}_0}
\def\Mttilde{\widetilde{M}_0}
\def\mttilde{\widetilde{m}_0}
\def\Ftbetabar{F_{0}\supbetabar}
\def\Fhat{\widehat{F}}
\def\Ftbetainv{F_{0}\supbetainv}
\def\Fthatbetainv{\Fhat_{0}\supbetainv}
\def\Fthatbeta{\Fhat_{0}\supbeta}
\def\Fthatbetahat{\Fhat_{0}\supbetahat}
\def\Ftbeta{F_{0}\supbeta}
\def\wbar{\bar{w}}
\def\what{\widehat{w}}
\def\whati{\what_i}
\def\wbari{\wbar_i}
\def\Ebb{\mathbb{E}}
\def\Pbb{\mathbb{P}}
\def\Ebbhat{\widehat{\Ebb}}
\def\Pbbhat{\widehat{\mathbb{P}}}
\def\Ebbt{\mathbb{E}_0}
\def\Pbbt{\mathbb{P}_0}
\def\Ebbs{\mathbb{E}_1}
\def\Pbbs{\mathbb{P}_1}
\def\sumin{\sum_{i=1}^n}
\def\sumiN{\sum_{i=1}^N}
\def\sumiNt{\sum_{i=1}^{N_t}}
\def\sumjNt{\sum_{j=1}^{N_t}}
\def\supSL{^{\scriptscriptstyle \sf SL}}
\def\supSTEAM{^{\scriptscriptstyle \sf STEAM}}
\def\supSTEAMstar{^{\scriptscriptstyle \sf STEAM*}}
\def\supSTEAMk{^{\scriptscriptstyle \sf STEAM,(k)}}
\def\wtilde{\widetilde{w}}
\def\cbar{\bar{c}}
\def\chat{\widehat{c}}
\def\subut{_{u_0}}
\def\subbeta{_{\scriptscriptstyle \bbeta}}
\def\subbetahat{_{\scriptscriptstyle \bbetahat}}
\def\subbetabar{_{\scriptscriptstyle \bbetabar}}
\def\supbetainv{^{\scriptscriptstyle \bbeta^{-1}}}
\def\supbeta{^{\scriptscriptstyle \bbeta}}
\def\supbetahat{^{\scriptscriptstyle \bbetahat}}
\def\supbetahatstar{^{\scriptscriptstyle \bbetahat^\ast}}
\def\supalphahat{^{\scriptscriptstyle \balphahat}}
\def\supalpha{^{\scriptscriptstyle \balpha}}
\def\supalphabar{^{\scriptscriptstyle \balphabar}}
\def\supbetabar{^{\scriptscriptstyle \bbetabar}}
\def\supbetahatnk{^{\scriptscriptstyle \bbetahat^{(-k)}}}
\def\bX{\bm X}
\def\bx{\bm x}
\def\IL{\mathcal I_L}
\def\IU{\mathcal I_U}
\def\IT{\mathcal I_T}
\def\pto{\overset{P}\to}
\def\Psc{\mathcal{P}}
\def\Psct{\mathcal{P}_0}
\def\Pscthat{{\widehat\Psc}_0}
\def\Ptbeta{\Psct\supbeta}
\def\Ptbetabar{\Psct\supbetabar}
\def\Psctbbetahat{\Psct\supbetahat}
\def\Pthatbeta{\Pscthat\supbeta}
\def\Pthatbetahat{\Pscthat\supbetahat}
\def\What{\widehat{W}}
\def\Wsc{\mathcal{W}}
\def\Wschat{\widehat{\Wsc}}
\def\trans{^{\scriptscriptstyle \sf T}}
\def\Gbb{\mathbb{G}}
\def\Hsc{\mathcal{H}}
\def\bB{\mathbf{B}}
\def\bC{\mathbf{C}}

\def\Ptbetabari{ {\mathcal P}_{0,i}\supbetabar}
\def\Ptbetabarj{ {\mathcal P}_{0,j}\supbetabar}
\def\Ptbetahati{ {\mathcal P}_{0,i}\supbetahat}
\def\Pthatbetahat{\widehat{\Psc}_{0}\supbetahat}
\def\Pthatbetahatstar{\widehat{\Psc}_{0}\supbetahatstar}
\def\Pthatbetahati{\widehat{\Psc}_{0,i}\supbetahat}
\def\Pthatbetahatstari{\widehat{\Psc}_{0,i}\supbetahatstar}
\def\Pthatbetahatnki{\widehat{\Psc}_{0,i}\supbetahatnk}
\def\Pthatbetabar{\widehat{\Psc}_{0}\supbetabar}
\def\Pthatbetabari{\widehat{\Psc}_{0,i}\supbetabar}
\def\Ptbetai{\Psc_{0,i}\supbeta}

\def\supL{^{\scriptscriptstyle \mathcal{L}}}
\def\supU{^{\scriptscriptstyle \mathcal{U}}}
\def\mbar{\bar{m}}
\def\pihat {\widehat \pi}
\def\what {\widehat w}

\newrefsection
\section*{Appendix}
In the Appendix, we outline theoretical derivations to establish the consistency and asymptotic normality of our proposed estimator. Let $\Pbb_s$ and $\Ebb_s$ denote the probability measure and expectations in the population $S = s$; let $\{\Pbbhat_s\supL, \Ebbhat_s\supL\}$ and $\{\Pbbhat_s\supU, \Ebbhat_s\supU\}$ denote the corresponding empirical measures in the labeled and unlabeled set; and let $\Gbb_n\supL = \nhalf(\Pbbhat_t\supL-\Pbb_t)$ and $\Gbb_N\supU = \Nthalf(\Pbbhat_t\supU - \Pbb_t)$.
As defined in the main text, let $w(\bX)=\Pbb_0(\bX)/\Pbb_1(\bX)=\{1-\pi(\bX)\}/\pi(\bX)\cdot P(S=1)/P(S=0)$, $\pi(\bX)=P(S=1|\bX)$. Let $w(\bX;\balpha,\bbeta)=\{1-\pi(\bX;\balpha,\bbeta)\}/\pi(\bX;\balpha,\bbeta)\cdot P(S=1)/P(S=0)$, $\pi(\bX;\balpha,\bbeta)=P(S=1\mid \balpha\trans\bX,\bbeta\trans\bX)$ and $\what(\bX;\balpha,\bbeta)=\{1-\pihat(\bX;\balpha,\bbeta)\}/\pihat(\bX;\balpha,\bbeta)$. Besides the assumptions made in Section 2.1, we also assume the following:
\begin{enumerate}
    \item \label{ass_comp} $\balpha$ and $\bbeta$ belongs to compact parameter spaces, and the covariates $\bX$ are bounded.
    \item \label{ass_cont} $\bbeta\trans \bX$ and $\balpha\trans \bX$ are continuous with continuously differentiable density functions. $P(Y=1|\bX)$ and $P(S=1|\bX)$ are twice continuously differentiable.
    \item \label{ass_h}
    The kernel $K$ is a symmetric, bounded and continuously differentiable kernel. We set the bandwidths $h_1$ and $h_2$ to $h_1 = O(N^{-\nu_1})$ and $h_2 = O(n^{-\nu_2})$ and $N \gtrsim n^{r_0}$ with $\nu_1 \in [1/6, 1/2)$, $\nu_2 \in (1/4, 1/2)$, $ \nu_1 < \half(1-\frac{1}{r_0})$, $r_0 > \max(4\nu_2, 1+2\nu_2) = 1+2\nu_2$. 
    \item \label{ass_dr} At least one of the following two conditions holds, (i) The weighting model is is correct, $w(\bX_i;\bar\alpha, \bar\beta) = w(\bm \bX_i)$. (ii) The outcome model is correct, $\Ebb(Y_i| \bar\balpha\trans  \bX_i ,\bar\bbeta\trans  \bX_i) = \mu(\bX_i)$.
\end{enumerate}

\begin{remark}
If we choose undersmoothing for outcome regression with $\nu_2 = 1/3$, then we need $r_0 > 5/3$ and can choose $\nu_1 < \half(1-\frac{1}{r_0}) < \frac{1}{5}$. 
\end{remark}

\subsection*{A. Consistency}

First, we establish the consistency of $\TPRhat\supSTEAM_0(c)$ when either the outcome model \eqref{eq:mu} or the density ratio model \eqref{eq:pi} is correctly specified. From the weak convergence of $\Pthatbeta(\bx) = \Fthatbeta(\bbeta\trans\bx)$ to $\Ptbeta(\bx)=\Ftbeta(\bbeta\trans\bx)  $ as a process in $\bx$ and $\bbeta$ \parencite{wellner2013weak}, 
\begin{equation}
\sup_{\bbeta,\bX} \left|\Pthatbeta(\bX) - \Ptbeta(\bX)\right| = O_p(\Nnhalf) \label{rate-phat}
\end{equation}
and $\widehat \Wsc(\bbeta,\bx) = \Nhalf\left\{ \Pthatbeta(\bx) - \Ptbeta(\bx) \right\}$
converges weakly as a process in $\bbeta$ and $\bx$. This implies stochastic continuity of $\Wschat(\bbeta,\bX)$ in $\bbeta$ and $\bX$. Thus
$\Wschat(\bbetahat,\bx) - \Wschat(\bbeta,\bx) \to 0$ in probability. 

With the convergence of $\balphahat, \bbetahat$ to $\bar{\balpha}, \bar{\bbeta}$, by \parencite{zou_adaptive_2009}, the uniform convergence rates of kernel estimators \parencite{ullah1999nonparametric},  we have 
\begin{alignat}{2}
& \max_i |\Pthatbetai - \Ptbetai| = O_p(\Nnhalf) , \quad \sup_{\bx, \bbeta}|\what(\bx; \balphahat,\bbeta) - w(\bx, \balphabar, \bbeta)| = O_p\left( \sqrt{\frac{\log N}{Nh_1^2}}\right) , 
\label{rate-phati-what-beta} \\
& \max_i |\Pthatbetahati - \Ptbetabari| = O_p(\nnhalf + \Nnhalf) , \quad \mbox{and} \quad \max_i|\what_i - \wbar_i| = O_p\left(\nnhalf +  \sqrt{\frac{\log N}{Nh_1^2}}\right) , 
\label{rate-phati-what} 
\end{alignat}
where $\wbar_i = w(\bX_i, \balphabar, \bbetabar)$. It then follows from the uniform consistency of the kernel estimators \parencite{ullah1999nonparametric} and arguments similar to those given in \parencite{gronsbell_semi_2018} that 
\begin{eqnarray*}
&&\mthat(q,\Pthatbetahat,\what(\cdot, \balphahat,\bbetahat)):= \frac{\ninv\sum_{i\in\mathcal I_L}K_{m,h_2}( \Pthatbetahati-q)Y_i \what(\bX_i, \balphahat,\bbetahat)}{\ninv\sum_{i\in\mathcal I_L}K_{m,h_2}(\Pthatbetahati-q)\what(\bX_i, \balphahat,\bbetahat)}\\
&\rightarrow_p  &\mbar_0(q) := \mbar_0(q, {\mathcal P}_0^{\bar\bbeta},w(\cdot,\balphabar,\bbetabar)) =
\frac{\Ebb_1\{Y_i \wbari K_{m,h_2}( \Ptbetabari-q)\} }{\Ebb_1\{\wbari K_{m,h_2}( \Ptbetabari-q)\}} ,
\end{eqnarray*}
where $\rightarrow_p$ means converges in probability.
Next, we will show that under assumption \ref{ass_dr}, i.e. if either model \eqref{eq:mu} or model \eqref{eq:pi} is correct, 
\begin{eqnarray}
 \mbar_0(q)  
&\rightarrow_p&  m_0(q) := m_0(q, {\mathcal P}_0^{\bar\bbeta}) = \frac{ m_0(q, {\mathcal P}_0^{\bar\bbeta})f_{{\mathcal P}_0^{\bar\bbeta}}(q)}{f_{{\mathcal P}_0^{\bar\beta}}(q)}, \label{limit}
\end{eqnarray}
where $m_0(q, {\mathcal P}_0^{\bar\beta})=\Ebbt(Y|{\mathcal P}_0^{\bar\beta} =q)$ and $f_{{\mathcal P}_0^{\bar\bbeta}}(\cdot)$ is the density function of ${\mathcal P}_0^{\bar\bbeta}$. If model \eqref{eq:pi} is valid, $\wbari = w(\bm \bX_i)$, we have that
\begin{eqnarray*}
\Ebbs\{  K_{m,h_2}(\Ptbetabari-q) Y_i \wbari \}&=& \Ebbt\{ K_{m,h_2}(\Ptbetabari-q) Y_i  \}
\rightarrow_p m_0(q, {\mathcal P}_0^{\bar\bbeta})f_{{\mathcal P}_0^{\bar\bbeta}}(q),\\
\Ebbs\{  K_{m,h_2}(\Ptbetabari-q)  \wbari \}&\rightarrow_p&f_{{\mathcal P}_0^{\bar\bbeta}}(q),
\end{eqnarray*}
so (\ref{limit}) follows.
On the other hand, if model \eqref{eq:mu} is correct, then $\Ebb(Y_i \mid \bX_i)  = g(\bar\bbeta\trans\bX_i)$ and $Y_i \perp S_i \mid \bar\bbeta\trans\bX_i$, and thus we have
\begin{eqnarray*}
\Ebbs\{  K_{m,h_2}(\Ptbetabari-q) Y_i \wbari \}&=&  \Ebbs\{  K_{m,h_2}(\Ptbetabari-q) g(\bbetabar\trans\bX_i) \wbari | \bar\balpha\trans  \bX_i ,\bar\bbeta\trans  \bX_i \}\\
&=& \Ebbt\{  K_{m,h_2}(\Ptbetabari-q) g(\bbetabar\trans\bX_i) \}
\rightarrow_p m_0(q, {\mathcal P}_0^{\bar\bbeta})f_{{\mathcal P}_0^{\bar\bbeta}}(q),\\
\Ebbs\{  K_{m,h_2}(\Ptbetabari-q)  \wbari \}&\rightarrow_p&f_{{\mathcal P}_0^{\bar\bbeta}}(q),
\end{eqnarray*}
which implies (\ref{limit}).
Combining the consistency of $\mthat(q,\Pthatbetahat,\what(\cdot, \balphahat,\bbetahat))$ to $m_0(q, {\mathcal P}_0^{\bbetabar})$ with the convergence of $\max_i|\Pthatbetahati-\Ptbetabari|\rightarrow_p 0$ , we have
$$
\TPRhat\supSTEAM_0(c) \equiv \frac{\sum_{i\in\mathcal I_T}  I(\Pthatbetahati \ge c) \mthat(\Pthatbetahati,\Pthatbetahat,\what(\cdot, \balphahat,\bbetahat))}{\sum_{i\in\mathcal I_T}  \mthat(\Pthatbetahati,\Pthatbetahat,\what(\cdot, \balphahat,\bbetahat))}   \rightarrow_p \TPR_0(c) \equiv \frac{\Pbb_0(\mathcal P_0^{\bar\beta}\ge c, Y=1)}{\Pbb_0(Y=1)}
$$
when either model \eqref{eq:mu} or \eqref{eq:pi} is correct.

\def\bphi{\boldsymbol{\phi}}

\subsection*{B. Asymptotic normality}
Next, we establish the asymptotic normality of $\nhalf\{\TPRhat\supSTEAM_0(c)-\TPR_0(c)\}$. From \textcite{zou_adaptive_2009}, we have the asymptotic normality and oracle properties for $\balphahat$ and $\bbetahat$ such that 
\begin{eqnarray*}
\Nhalf ({\balphahat}-\balphabar)&=& \Nnhalf\sum_{i=1}^N \bphi_i\supalphabar+o_p(1),\\
\nhalf (\bbetahat-\bbetabar)&=& \nnhalf\sum_{i=1}^{n} \bphi_i\supbetabar+o_p(1),
\end{eqnarray*}
where $\bphi_i\supalphabar, \bphi_i\supbetabar$ are respective influence functions. Due to oracle properties, the elements of $\bphi_i\supalphabar$ and  $\bphi_i\supbetabar$ corresponding to the zero components of $\balphabar$ and $\bbetabar$ also take value zero. Let
$$
\TPRhat\supSTEAM_0(c) = \frac{N_t^{-1}\sum_{i\in\mathcal I_T}  I(\Pthatbetahati \ge c) \mthat(\Pthatbetahati,\Pthatbetahat,\what(\cdot, \balphahat,\bbetahat))}{N_t^{-1}\sum_{i\in\mathcal I_T}  \mthat(\Pthatbetahati,\hat{\mathcal P}_{\bbetahat},\what(\cdot, \balphahat,\bbetahat))}  =\frac{\widehat{\xi}(c, \bbetahat)}{\widehat{\xi}(0, \bbetahat)} ,
$$
 $\xi\supalpha(c,\bbeta)=E\{I(\Ptbetai \ge c) m_0(\Ptbetai,\Ptbeta,w(\cdot, \balpha, \bbeta) \}$, and $\xi(c, \bbeta) = \xi\supalphabar(c,\bbeta)$. We write
\begin{eqnarray*}
 \nhalf\{\widehat{\xi}(c, \bbetahat) - \xi(c,\bbetabar)\}
&=&\nhalf N_t^{-1}\sum_{i\in\mathcal I_T}  I(\Pthatbetahati \ge c) \mthat(\Pthatbetahati,\Pthatbetahat,\what(\cdot, \balphahat,\bbetahat))-\nhalf \xi(c,\bbetabar)\\
&=& \What_1(\bbetahat) + \What_2+ \What_3+\What_4 
\end{eqnarray*}
where
\begin{eqnarray*}
\What_1(\bbeta) 
&=& \nhalf N_t^{-1}\sum_{i\in\mathcal I_T}  \left\{ I(\Pthatbetai \ge c) \Mthat(\Pthatbetai; \Pthatbeta, \balphahat)-  I(\Ptbetai\ge c) \Mthat(\Ptbetai;\Ptbeta, \balphahat) \right\}\\
\What_2 &=& \nhalf N_t^{-1}\sum_{i\in\mathcal I_T}    \left\{I(\Ptbetahati \ge c)\Mthat\supbetahat(\Ptbetahati,\Ptbetahat, \balphahat) -   I(\Ptbetabari \ge c)\Mttilde\supbetabar(\Ptbetabari,\Ptbetabar, \balphabar) \right\}\\
\What_3 &=&\nhalf N_t^{-1}\sum_{i\in\mathcal I_T}  I( \Ptbetabari \ge c) \left\{ \mthat(\Ptbetabari,\Ptbetabar,w(\cdot, \balphabar,\bbetabar))- m_0(\Ptbetabari,\Ptbetabar )  \right\} \\
\What_4 &=&\nhalf N_t^{-1}\sum_{i\in\mathcal I_T}  \left\{ I(\Ptbetabari \ge c)  m_0(\Ptbetabari,\Ptbetabar ) - \xi(c,\bbetabar) \right\} , 
\end{eqnarray*}
$\Mthat\supbeta(u;\Psc, \alpha) = \mthat(u,\Psc,\what(\cdot, \balpha,\bbeta))$ and $\Mttilde\supbeta(u;\Psc, \alpha) = \mthat(u,\Psc,w(\cdot, \balpha,\bbeta))$. It is straightforward to see that 
\begin{eqnarray*}
W_4&=&\nhalf \left\{N_t^{-1}\sum_{i\in\mathcal I_T}  I(\Ptbetabari \ge c)  m_0(\Ptbetabari,\Ptbetabar) -  \xi(c,\bbetabar)\right\}=O_p(\nhalf \Nnhalf)=o_p(1) .
\end{eqnarray*}
We next approximate each of the remaining three terms.

To approximate $\What_1(\bbetahat)$, we write  $\What_1(\bbeta) = \What_{11}(\bbeta) + \What_{12}(\bbeta)$ with
\begin{eqnarray*}
\What_{11}(\bbeta) &=&\nhalf N_t^{-1}\sum_{i\in\mathcal I_T}  \{I(\Pthatbetai \ge c)-I(\Ptbetai \ge c)\}
\Mthat(\Pthatbetai; \Pthatbeta, \balphahat)\\
\What_{12}(\bbeta)&=& \nhalf N_t^{-1}\sum_{i\in\mathcal I_T}  I(\Ptbetai \ge c) \{\Mthat(\Pthatbetai; \Pthatbeta, \balphahat)-\Mthat(\Ptbetai; \Ptbeta, \balphahat)\}
\end{eqnarray*}
For $ \What_{11}(\bbeta)$, we write
\begin{eqnarray*}
\What_{11}(\bbeta) & = & \nhalf\Ntnhalf \int_c^{1} 
\Mthat(u; \Pthatbeta, \balphahat) \Wschat_\Psc(du, \bbeta)
\end{eqnarray*}
where $\Wschat_\Psc(u, \bbeta) = \Ntnhalf \sum_{i=1}^{N_t}\{I(\Pthatbetai \le c) - I(\Ptbetai \le c)\}$ can be shown as convergent to a Gaussian process in $u$ and $\bbeta$. It then follows similar arguments given in Appendix A of \textcite{gronsbell_semi_2018}, we have 
\begin{align*}
& \nhalf \sup_{q, \bbeta} \left| \ninv \sumin K_{m,h_2}(\Pthatbetai - q)Y_i \what_iY_i^a - \ninv \sumin K_{m,h_2}(\Ptbetai - q)Y_i \wbar_iY_i^a  \right| \\
\lesssim & \sqrt{\frac{n\log N}{Nh_1^2}}  + \left| \int K_{m,h_2}(v-q) d \Gbb_n\supL\left[ I\{\bbeta\trans\bx \le \Fthatbetainv(v)\} y^a\wbar\supbeta - I\{\bbeta\trans\bx \le \Ftbetainv(v)\} y^a\wbar\supbeta\right] \right| \\
\lesssim & O_p\left(\sqrt{\frac{n\log N}{Nh_1^2}}\right)  +
h_2^{-1}\| \Gbb_n\supL\|_{\Hsc_\delta} \lesssim O_p\left\{ (\nhalf N^{-\half + \nu_1}  + n^{\nu_2} N^{-\frac{1}{4}})\log(N) \right\} \\
\lesssim & O_p\left[\{N^{-\half(1-1/r_0- 2\nu_1)}+N^{-\frac{1}{4}+\nu_2/r_0}\}\log(N) \right]
\end{align*}
where $\wbar\supbeta = w(\bx, \balphabar, \bbeta)$,
$\Hsc_\delta =\{I(\bbeta\trans\bx \le s)y^a\wbar\supbeta - I(\bbeta\trans\bx \le s')y^a\wbar\supbeta: \bbeta, |s-s'| < \delta \}$ and $\|\Gbb_n\supL\|_{\Hsc_{\delta}} \lesssim O_p\{N^{-1/4}\log(N)\}$.
For $\What_{12}(\bbeta)$, we note that 
$\max_{i,j}|(\Pthatbetai - \Pthatbetaj) - (\Ptbetai - \Ptbetaj)| = O_p(\Ntnhalf)$. By a Taylor series expansion, we have 
\begin{align*}
\sup_{\bbeta}\left|\What_{12}(\bbeta)\right| & \le \nhalf\sup_{\bbeta}\left| N_t^{-1}\sum_{i\in\mathcal I_T}  |\Mthat(\Pthatbetai; \Pthatbeta, \balphahat)-\Mthat(\Ptbetai; \Ptbeta, \balphahat)| \right| \\
& \lesssim O_p(h_2^{-1} \nhalf \Nnhalf) = 
O_p\left(N^{-\frac{1- 1/r_0 - 2\nu_2/r_0}{2}}\right)
\end{align*}
It follows from Assumption 3 that $|\What_1(\bbetahat)| \le  \sup_{\bbeta}|\What_{11}(\bbeta)|+\sup_{\bbeta}|\What_{12}(\bbeta)|
= o_p(1)$.

For the second term, we write $\What_2 = \What_{21}+\What_{22}$, where
\begin{align*}
\What_{21} & = \nhalf N_t^{-1}\sum_{i\in\mathcal I_T}   I(\Ptbetahati \ge c) \left\{\Mthat\supbetahat(\Ptbetahati,\Ptbetahat, \balphahat) -   \Mttilde\supbetahat(\Ptbetahati,\Ptbetahat, \balphabar) \right\}\\ 
\What_{22} & = \nhalf N_t^{-1}\sum_{i\in\mathcal I_T}    \left\{I(\Ptbetahati \ge c)\Mttilde\supbetahat(\Ptbetahati,\Ptbetahat, \balphabar) -   I(\Ptbetabari \ge c)\Mttilde\supbetabar(\Ptbetabari,\Ptbetabar, \balphabar) \right\} 
\end{align*}
From (\ref{rate-phati-what-beta}), we have $|\What_{21}| \lesssim  O_p\left(\sqrt{\frac{n\log N}{Nh_1^2}}\right)=o_p(1)$. Next, we expand $\What_{22}$ as $\What_{22} = \What_{22,1}+\What_{22,2}+\What_{22,3}$, where
\begin{eqnarray*}
\What_{22,1}
&=&\nhalf \int_c^{1} \left\{\mthat(u, \Ptbetahat, w(\cdot,\balphabar,\bbetahat)) - m_0(u, \Ptbetabar, w(\cdot,\balphabar,\bbetabar)) \right\}d \left\{\Fschat\supbetahat(u) -  \Fsc\supbetabar(u)\right\}  \\
\What_{22,2}&=&\nhalf \int_c^{1} m_0(u, \Ptbetabar, w(\cdot,\balphabar,\bbetabar)) d \left\{\Fschat\supbetahat(u) -  \Fsc\supbetabar(u)\right\}  \\
\What_{22,3}&=& \nhalf \int_c^{1} \left\{\mthat(u, \Ptbetahat, w(\cdot,\balphabar,\bbetahat)) - \mthat(u, \Ptbetabar, w(\cdot,\balphabar,\bbetabar)) \right\} d \Fsc\supbetabar(u)
\end{eqnarray*}
where $\Fschat\supbeta(c) = \Ntinv\sumiNt I(\Ptbetai \le c)$ and
$\Fsc\supbeta(c) = \Pbbt(\Ptbetai\le c)$. From Lemma A.2 and A.3 of \textcite{chakrabortty2018efficient}, we have $\What_{22,1} = o_p(1)$. Since $\Gbb_N\supU\{I(\Ptbetai \le c)\}$ converges weakly to a Gaussian process in $\bbeta$ and $c$, we have 
$$
\What_{22,2}=\nhalf (\bbetahat-\bbetabar)\trans\int_c^{\infty} m_0(u, \Ptbetabar, w(\cdot,\balphabar,\bbetabar)) d \Fscdot\supbetabar(u)  +o_p(1) 
$$
where $\Fscdot\supbeta(u) = \partial \Fsc\supbeta(u)/\partial \bbeta$. 
For $\What_{22,3}$, we take a Taylor series expansion to obtain 
$$
\What_{22,3}= \nhalf(\bbetahat-\bbetabar)\trans \int_c^{\infty} \dot{m}_0\supbetabar(u,\Ptbetabar, w(\cdot,\balphabar, \bbetabar) d \Fsc\supbetabar(u) + o_p(1)
$$
where $\dot{m}_0\supbeta(u, \Ptbeta, w(\cdot, \balphabar, \bbeta) = 
\partial m_0(u, \Ptbeta, w(\cdot, \balphabar, \bbeta)) / \partial \bbeta$. This, together with other terms associated with $\What_2$, we have
$$
\What_{2} = \nhalf(\bbetahat-\bbetabar)\trans\bC_2 + o_p(1).
$$
where $\bC_2 = \int_c^{\infty} m_0(u, \Ptbetabar, w(\cdot,\balphabar,\bbetabar)) d \Fscdot\supbetabar(u) + \int_c^{\infty} \dot{m}_0\supbetabar(u,\Ptbetabar, w(\cdot,\balphabar, \bbetabar) d \Fsc\supbetabar(u)$.

For $\What_3$, we again use  Lemma A.2 of \textcite{chakrabortty2018efficient} and write 
\begin{eqnarray*}
 \What_3&=&\nhalf N_t^{-1}\sum_{i\in\mathcal I_T}  I(\Ptbetabari \ge c)\{ \mthat(\Ptbetabari,\Ptbetabar,w(\cdot, \balphabar,\bbetabar))- \mbar_0(\Ptbetabari) \}\\
&=& \nhalf \int_c^1   \frac{n^{-1}\sum_{j\in\mathcal I_L}K_{m,h_2}( \Ptbetabarj-s)\wbar_j \{Y_j -m_0(s,\Ptbetabar ) \}}{n^{-1}\sum_{j\in\mathcal I_L}K_{m,h_2}(\Ptbetabarj-s)\wbar_j}d \Fschat\supbetabar(s) +o_p(1)\\
&=& \nnhalf \sum_{j\in\mathcal I_L} \int_c^1   K_{m,h_2}( \Ptbetabarj-s)\wbar_j \{Y_j -m_0(s,\Ptbetabar ) \}d\Fsc \supbetabar(s) +o_p(1)\\
&=& \nnhalf \sum_{j\in\mathcal I_L} \wbar_jI(\Ptbetabarj > c) \{ Y_j  -m_0(\Ptbetabarj,\Ptbetabar )\} +o_p(1) .
\end{eqnarray*}

Combining expansions for $\What_2$ and $\What_3$, we have
$$\nhalf\{\xihat(c,\bbetahat)-\xi(c,\bbetabar)\}=\nnhalf \sum_{j\in\mathcal I_L} \left[ \wbar_jI(\Ptbetabarj > c)  \{Y_j  -\xi(c,\bbetabar )\} +  \bC_2\trans\bphi_i\supbeta \right]+ o_p(1),$$
which converges in distribution to a normal. The asymptotic normality of $\nhalf\{\TPRhat\supSTEAM_0(c)-\TPR_0(c)\}$ thus follows, and similarly, we can establish the other accuracy parameters. 

\subsection*{C. Cross-validation procedure}
It is important to note that the proposed estimators are subject to overfitting because the data in $\mathcal L$ is used to estimate both the risk score for classification and the conditional risk functions $m(\cdot)$, which are used for estimation of the accuracy parameters. To reduce the over-fitting bias in the proposed estimators, we propose a $K$-fold cross-validation procedure. CV in this setting involves partitioning $\mathcal L$ for the estimation of $\bbeta$ and $m_0(\cdot)$, whereas use all $\mathcal T$ to estimate the performance measures. Denote each fold of $\mathcal L$ as $\mathcal L_k$ and the corresponding indices as $\mathcal I_k$ for $k = 1,\dots , K$. For a given $k$, we fit the regression model with $\mathcal L\backslash \mathcal L_k$ to obtain an estimator for $\bbeta$ denoted as $\bbetahat^{(-k)}$. The observations in the left-out set $\mathcal L_k$ are used to estimate $m_0(\cdot)$ with the local constant smoother as
$$
\widehat m_{0,CV}\supbetahat(q) = \frac{\sum_k\sum_{i\in\mathcal I_k}K_{m,h_2}( \Pthatbetahatnki-q)\what_i^{(-k)}Y_i}{\sum_k\sum_{i\in\mathcal I_k}K_{m,h_2}(\Pthatbetahatnki-q)\what_i^{(-k)}} 
$$
where
$$
   \what_i^{(-k)} = \frac{1-\pihat^{(-k)}_i}{\pihat^{(-k)}_i} 
$$
with
$$
    \pihat ^{(-k)}_i  = \pihat^{(-k)}(\bX_i; \balphahat, \bbetahat^{(-k)}) = \frac{\sum_{j} K_h\{(\balphahat, \bbetahat^{(-k)})\trans  (\bX_j-\bX_i)\}S_i}{\sum_{j}  K_h\{(\balphahat, \bbetahat^{(-k)})\trans  (\bX_j-\bX_i)\}} 
$$
where
$$
\bbetahat^{(-k)} = \argmax_{\bm\beta} \left\{n^{-1} \sum_{i\in\IL\backslash \mathcal I_k} \ell(\bbeta;Y_i,\bX_i) - \lambda_{\mu,n}  \|(\bbeta_{-1}/\bbetatilde^\ast_{-1})^\gamma\|_1 \right\}
$$
We then estimate $\TPR_0(c)$ as
$$
    \TPRhat\supSTEAM_{0,CV}(c) = \frac{\sum_{i\in\IT}  I(\Pthatbetahat \ge c) \widehat m_{0,CV}(\Pthatbetahat)}{\sum_{i\in\IT}  \widehat m_{0,CV}(\Pthatbetahat)}  
$$
Similarly, we may construct CV estimators for $\FPR_0(c)$, $\ROC_0(c)$, $\PPV_0(c)$ and $\NPV_0(c)$, respective denoted as $ \FPRhat\supSTEAM_{0,CV}(c)$, $ \ROChat\supSTEAM_{0,CV}(c)$, $ \PPVhat\supSTEAM_{0,CV}(c)$, $ \NPVhat\supSTEAM_{0,CV}(c)$.

\subsection*{D. Estimated ROC curves for RA phenotyping model}
We present the estimated ROC curves along with their respective $95\%$ point-wise confidence intervals 
\clearpage
\begin{figure}[ht]
\centering
\includegraphics[width=0.9\textwidth]{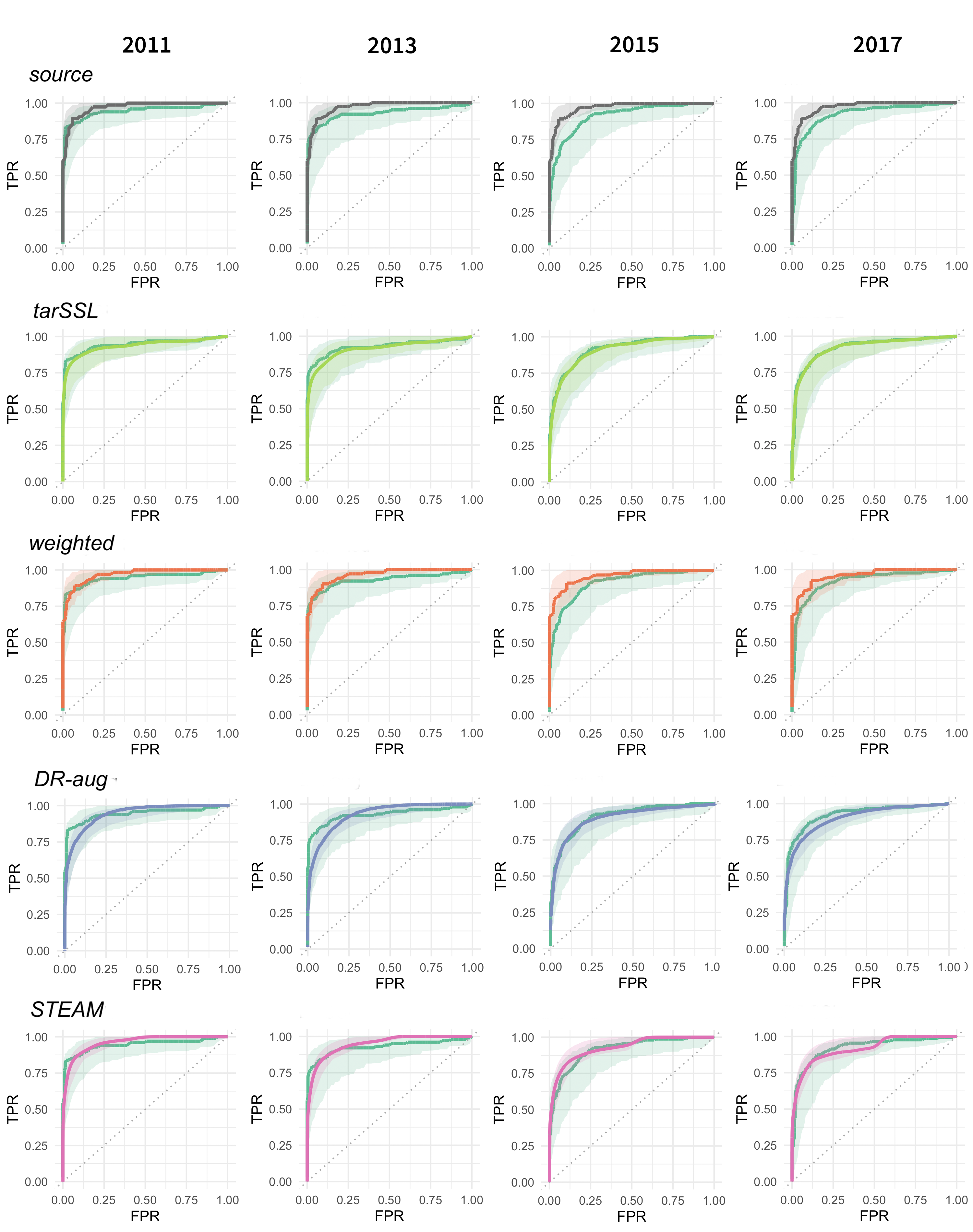}
\captionsetup{labelformat=empty}
\caption*{Figure S1: Estimated ROC curves along with corresponding 95\% confidence bands of \textit{ALASSO-2009} on dataset in each year based on \textit{naive} (grey), \textit{weighted} (orange), \textit{DR-aug} (purple) and \textit{STEAM} (pink) methods, compared with \textit{target\_labeled} (green).}
\addtocounter{figure}{-1} 
\end{figure}
\clearpage

\printbibliography

\end{document}